\documentclass[aps,prb,twocolumn,showpacs,10pt,floatfix,superscriptaddress,floatfix]{revtex4-1}
\usepackage[dvips]{graphics}
\usepackage{makecell}
\usepackage{boldline}
\usepackage{graphicx}
\usepackage[utf8]{inputenc}
\usepackage{amssymb}
\usepackage{epstopdf}
\usepackage{bm}
\usepackage{dcolumn,multirow}
\usepackage{epsfig}
\usepackage{latexsym}
\usepackage{amsmath,mathtools}
\usepackage{color}
\usepackage{array}
\usepackage{enumerate,dsfont}
\newcommand{\bZ}{\mathbb{Z}}

\def\XXint#1#2#3{{\setbox0=\hbox{$#1{#2#3}{\int}$ }
\vcenter{\hbox{$#2#3$ }}\kern-.5\wd0}}

\newcommand{\ket}[1]{\left |\mbox{$#1$}\right\rangle}
\newcommand{\bra}[1]{\left\langle\mbox{$#1$}\right |}

\newcommand{\vctr}[1]{{\bf {#1}}}
\newcommand{\vk}{\vctr{k}}

\newcommand{\cH}[1]{\mathcal{H}_{#1}}
\newcommand{\cR}[1]{\mathcal{M}_{#1}}
\newcommand{\ccR}[1]{\mathcal{R}_{#1}}

\makeatletter
\newcommand*{\rom}[1]{\expandafter\@slowromancap\romannumeral #1@}
\makeatother

\begin{document}

\title{Bott periodicity for the topological classification of gapped states of matter with reflection symmetry}

\author{Luka Trifunovic and Piet Brouwer}
\affiliation{Dahlem Center for Complex Quantum Systems and Physics Department, Freie Universit\"at Berlin, Arnimallee 14, 14195 Berlin, Germany}
\date{\today}

\begin{abstract}
Using a dimensional reduction scheme based on scattering theory, we show that the classification tables for topological insulators and superconductors with reflection symmetry can be organized in two period-two and four period-eight cycles, similar to the Bott periodicity found for topological insulators and superconductors without spatial symmetries. With the help of the dimensional reduction scheme the classification in arbitrary dimensions $d \ge 1$ can be obtained from the classification in one dimension, for which we present a derivation based on relative homotopy groups and exact sequences to classify one-dimensional insulators and superconductors with reflection symmetry. The resulting classification is fully consistent with a comprehensive classification obtained recently by Shiozaki and Sato [Phys.\ Rev.\ B {\bf 90}, 165114 (2014)]. The use of a scattering-matrix inspired method allows us to address the second descendant $\bZ_2$ phase, for which the topological nontrivial phase was previously reported to be vulnerable to perturbations that break translation symmetry.
\end{abstract}

\maketitle
\section{Introduction}

The non-spatial symmetries time-reversal symmetry, particle-hole symmetry,
chiral symmetry, and their combinations can be organized in two ``complex'' and
eight ``real'' Altland-Zirnbauer symmetry classes.~\cite{altland1997} For each
of these classes the classification of the topological insulating
(non-interacting) phases of matter was recently
obtained.~\cite{PhysRevB.78.195125,doi:10.1063/1.3149495,1751-8121-44-4-045001,PhysRevB.85.085103,doi:10.1142/S0219887812500235,kennedy2016}
The classification has a cyclic dependence on the dimensionality, with a
period-two sequence for the complex classes and a period-eight sequence for the
real classes. This cyclic structure is known as ``Bott periodicity'' or ``Bott
clock'', and is firmly embedded in the mathematics of algebraic
topology~\cite{bott1959} and, in particular the eight ``real'' classes, in ``K
theory''.\cite{atiyah1966,karoubi2005}

An appealing physical construction that reproduces the period-two and
period-eight cyclic structure was suggested recently by Fulga and
coworkers.\cite{PhysRevB.85.165409} Fulga {\em et al.} show that the reflection
matrix $r_d$ of a gapped half-infinite system in $d$ dimensions with
Hamiltonian $H_d$ can be naturally interpreted as the Hamiltonian $H_{d-1}$ of
a gapped system in $d-1$ dimensions, but with a different symmetry, which
precisely follows the period-two and period-eight Bott clocks of the complex
and real Altland-Zirnbauer classes, respectively. Since topological
classification reflection matrix $r_d$ essentially amounts to a classification
of the {\em boundary} of the $d$-dimensional Hamiltonian $H_d$, the assumption
of bulk-boundary correspondence --- a topologically nontrivial bulk is
accompanied with gapless boundary states in a unique way --- then immediately
gives the Bott periodicity for non-interacting gapped phases of matter.

The concept of topological phases has been extended to include
(non-interacting) topological phases that are protected by a crystalline
symmetry, in addition to the non-spatial symmetries of the Altland-Zirnbauer
classification. The additional spatial symmetries give rise to gapless states
at the
boundary,\cite{PhysRevLett.89.077002,PhysRevB.73.214502,PhysRevB.78.045426,PhysRevB.81.134515,PhysRevB.83.064505,PhysRevB.85.165409,PhysRevB.83.224511,PhysRevB.84.060504,1367-2630-12-6-065004,PhysRevB.83.245132,PhysRevLett.105.097001}
provided the boundary is invariant under the symmetry operation. Although the
topological crystalline phases rely on spatial symmetries, many topological
phases associated with crystalline symmetries are robust to disorder that
preserves these symmetries on the
average.~\cite{PhysRevB.89.155424,PhysRevLett.109.246605,PhysRevB.86.045102,PhysRevLett.108.076804,PhysRevB.89.035117,PhysRevB.89.155315}

Because of the large number of possible spatial symmetries, especially in
higher dimensions, the task of classifying crystalline topological phases is a
formidable one. Although a comprehensive classification similar to that of the
non-spatial symmetries is still lacking, considerable progress has been made.
On the one hand, this includes complete classifications for all crystalline
symmetries, but restricted to a single Altland-Zirnbauer class in two or three
dimensions.~\cite{Slager2013,2016arXiv161202007K,PhysRevB.86.115112,PhysRevB.87.035119,PhysRevB.90.235141,PhysRevLett.113.116403,PhysRevLett.111.056403,PhysRevB.93.045429,PhysRevB.88.085110,PhysRevLett.111.047006,PhysRevB.89.224503}
On the other hand, there are classifications for all Altland-Zirnbauer classes
and all spatial dimensions, but for a restricted set of crystalline symmetry
operations, such as inversion symmetry~\cite{lu2014,2014arXiv1403.5558L} or
reflection
symmetry.~\cite{PhysRevB.88.075142,PhysRevB.88.125129,PhysRevB.90.165114} The
latter approach was found to yield period-two and period-eight dependencies on
the dimension $d$ known from the classification without crystalline
symmetries,~\cite{lu2014,PhysRevB.88.075142} including, in some cases, cyclic
structures reminiscent of the ``Bott
clock''.~\cite{PhysRevB.88.125129,PhysRevB.90.165114}

Reflection symmetry is one of most often considered crystalline symmetries, and
reflection-symmetric materials were the first experimental realizations of
crystalline topological insulators.~\cite{Hsieh2012,Tanaka2012,Hasan2012} A
complete classification of reflection-symmetric topological insulators and
superconductors was reported by Chiu {\em et al.}, employing the method of
minimal Dirac Hamiltonians.~\cite{PhysRevB.88.075142} Morimoto and
Furusaki,~\cite{PhysRevB.88.125129} using an approach based on Clifford
algebras,~\cite{doi:10.1063/1.3149495} showed that the topological classes with
reflection symmetry can be organized in period-two and period-eight cyclic
structure, although for some sequences the cycles involve increasing the
dimension $d$ in steps of two, rather than in unit steps, as in the case of the
standard Bott periodicity. Shiozaki and Sato generalized these results to all
order-two unitary and antiunitary crystalline
symmetries~\cite{PhysRevB.90.165114} and corrected some entries in the
classification table obtained
previously.~\cite{PhysRevB.88.075142,PhysRevB.88.125129}

In this paper we show that the reflection-matrix-based dimensional reduction
scheme of Fulga {\em et al.} can also be applied to reflection-symmetric
topological insulators. Using the scheme of Fulga {\em et al.} naturally leads
us to consider a ``chiral reflection'' operation, such that, once the chiral
reflection-symmetric gapped Hamiltonians are included, all symmetry
combinations can be grouped in period-two and period-eight cyclic sequences,
for which the dimension $d$ is increased in unit steps. Our results are in
complete agreement with the classification obtained by Shiozaki and Sato,
\cite{PhysRevB.90.165114} who used the Hamiltonian dimensional reduction scheme
of Teo and Kane\cite{PhysRevB.82.115120} to obtain relations between the
corresponding K groups. 

Whereas the reflection-matrix-based dimensional reduction scheme allows one to
obtain the classification for arbitrary dimension from the classification at
$d=0$ in the absence of spatial symmetries,\cite{PhysRevB.85.165409} in the
presence of reflection symmetry this procedure ends already for dimension
$d=1$, since a one-dimensional system has no ``boundary'' that is mapped onto
itself by reflection. To make this article self-contained and to provide an
alternative to the existing classification
schemes,\cite{PhysRevB.88.075142,PhysRevB.88.125129,PhysRevB.90.165114} we here
present a classification of one-dimensional reflection-symmetric topological
insulators based on relative homotopy groups and exact sequences, following the
approach taken by Turner {\em et al.} in their classification of
inversion-symmetric topological insulators.\cite{PhysRevB.85.165120} In
combination with the reflection-matrix-based reduction scheme, the $d=1$
classification gives a complete classification for reflection-symmetric
topological insulators in all dimensions $d \ge 1$. An additional advantage of
the approach of Turner {\em et al.} is that it gives explicit expressions for
topological invariants and for the generators of the topological classes (many
examples are given in App.\ \ref{app:exactsec}). 

Our approach allows us to address an issue related to stability of the
topological phase of the second descendant of $\bZ_2$ in the classes where
reflection symmetry anticommutes with non-spatial symmetries. Chiu {\em et al.}
and Morimoto and Furusaki argued that the topological $\bZ_2$ index cannot be
defined in these cases and that an eventual topologically non-trivial phase is
always ``weak'', {\em i.e.}, it is instable to perturbations that break the
lattice translation symmetry.\cite{PhysRevB.88.075142,PhysRevB.88.125129} While
Shiozaki and Sato left open the possibility of a ``subtle instability'' to
translation-symmetry-breaking perturbations, they insisted that the topological
invariant is a ``strong'' one.\cite{PhysRevB.90.165114} Having the explicit
form of the topological invariant at our disposal, we can confirm that it is
invariant under a redefinition of the unit cell. Moreover, since our
reflection-matrix based approach effectively classifies the boundary of the
insulator, we can show explicitly that a nonzero topological invariant implies
the existence of a topologically protected boundary state. We find no signs of
the instability reported in Refs.~\onlinecite{PhysRevB.88.075142,PhysRevB.88.125129}.

This article is organized as follows: In Sec.~\ref{sec:reflection} we review
the constraints that reflection symmetry poses on the Hamiltonian $H_d$ of a
gapped system in $d$ dimensions. In Sec.\ \ref{sec:methods} we review the
reflection-matrix-based method of dimensional reduction originally proposed by
Fulga {\em et al.}\cite{PhysRevB.85.165409} and we show how the method can be
generalized to reflection-symmetric topological insulators. The topological
classification of one-dimensional topological insulators with reflection
symmetry using the method of relative homotopy groups and exact sequences is
given in Sec.~\ref{sec:exactseq}. We discuss the controversial
second-descendant $\bZ_2$ phase in Sec.~\ref{sec:Z2}. We conclude with a brief
summary in Sec.~\ref{sec:summary}. Four appendices contain details of the
dimensionless reduction scheme, an extension of the $d=1$ classification to
higher dimensions ({\em i.e.}, without the assumption of bulk-boundary
correspondence, which underlies the reflection-matrix-based dimensional
reduction scheme), explicit examples for topological invariants of
one-dimensional reflection-symmetric topological insulators, and supporting
details for the second-descendant $\bZ_2$ phase.

\section{Symmetries}\label{sec:reflection}

We consider a Hamiltonian $H_d(\vk)$ in $d$ dimensions, with $\vk =
(k_1,k_2,\ldots,k_d)$. For definiteness we take the reflection plane to be
perpendicular to the $d$th unit vector, so that reflection maps the wavevector
$\vk$ to $R \vk = (k_1,k_2,\ldots,k_{d-1},-k_{d})$. Reflection also affects the
basis states in the unit cell, so that for the Hamiltonian $H_d(\vk)$
reflection symmetry takes the form
\begin{equation}
  H_d(\vk) = U_{\cal R}^{\dagger} H_d(R \vk) U_{\cal R},
\end{equation}
with $U_{\cal R}$ a $\vk$-independent unitary matrix. We require $U_{\cal R}^2 = 1$ to fix the phase freedom in the definition of $U_{\cal R}$.

The reflection symmetry exists possibly in combination with time-reversal (${\cal T}$), particle-hole (${\cal P}$), and/or chiral (${\cal C}$) symmetries. These
symmetries take the form
\begin{align}
  H_d(\vk) &= U_{\cal T}^{\dagger} H_d(-\vk)^* U_{\cal T}, \\
  H_d(\vk) &= - U_{\cal P}^{\dagger} H_d(-\vk)^* U_{\cal P}, \\
  H_d(\vk) &= - U_{\cal C}^{\dagger} H_d(\vk) U_{\cal C},
  \label{eq:chiralsymmetry}
\end{align}
where $U_{\cal T}$, $U_{\cal P}$, and $U_{\cal C}$ are $\vk$-independent
unitary matrices. If time-reversal symmetry and particle-hole symmetry are both
present, $U_{\cal C} = U_{\cal P} U_{\cal T}^*$. Further, the unitary matrices
$U_{\cal T}$, $U_{\cal P}$, and $U_{\cal C}$ satisfy $U_{\cal T} U_{\cal T}^* =
{\cal T}^2$, $U_{\cal P} U_{\cal P}^* = {\cal P}^2$, $U_{\cal C}^2 = 1$, and
$U_{\cal P} U_{\cal T}^* = {\cal T}^2 {\cal P}^2 U_{\cal T} U_{\cal P}^*$. [The condition $U_{\cal C}^2 = 1$ is not fundamental, since multiplication of $U_{\cal C}$ with a phase factor results in the same chiral symmetry relation (\ref{eq:chiralsymmetry}). We will use this condition to fix signs in intermediate expressions for the general derivation of the Bott clock from scattering theory, but not in the discussion of examples for specific symmetry classes.] 

The three non-spatial symmetry operations ${\cal T}$, ${\cal P}$, and ${\cal C}$ define the ten Altland-Zirnbauer classes.\cite{altland1997} The two ``complex'' classes have no symmetries linking $H$ to $H^*$; the remaining eight ``real'' classes have time-reversal symmetry, particle-hole symmetry, or both. Following common practice in the field, we 
use 
Cartan labels to refer to the corresponding symmetry classes, see Table \ref{tab:classes}.

\begin{table}
  \begin{tabular}{|c|c||c|c|c||c|c|}
    \hline 
    Class & Cartan & ${\cal T}$ & ${\cal P}$ & ${\cal C}$ & $\pi_0$ & $\pi_1$  \\\hline
    $\mathcal{C}_0$ & A & 0 & 0 & 0 &  $\bZ$ & 0 \\\hline
    $\mathcal{C}_1$ & AIII & 0 & 0 & 1  & 0 & $\bZ$ \\\hline \hline
    $\mathcal{R}_0$ & AI & 1 & 0 & 0 & $\bZ$ & $\bZ_2$ \\ \hline
    $\mathcal{R}_1$ & BDI & 1 & 1 & 1  & $\bZ_2$ & $\bZ_2$ \\ \hline
    $\mathcal{R}_2$ &  D & 0 & 1 & 0 & $\bZ_2$ & 0 \\ \hline
    $\mathcal{R}_3$ &  DIII & -1 & 1 & 1 & 0 & $2\bZ$ \\ \hline
    $\mathcal{R}_4$ &  AII & -1 & 0 & 0 & $2\bZ$ & 0 \\ \hline
    $\mathcal{R}_5$ &  CII & -1 & -1 & 1 & 0 & 0 \\ \hline
    $\mathcal{R}_6$ &  C & 0 & -1 & 0 & 0 & 0 \\ \hline
    $\mathcal{R}_7$ &  CI & 1 & -1 & 1 & 0 & $\bZ$ \\ \hline
  \end{tabular}
\caption{\label{tab:classes} The ten Altland-Zirnbauer classes are defined according to the presence or absence of time-reversal (${\cal T}$), particle-hole symmetry (${\cal P}$), and chiral symmetry (${\cal C}$). A nonzero entry indicates the square of the antiunitary symmetry operations ${\cal T}$ or ${\cal P}$.}
\end{table}

How the presence of a reflection symmetry affects the topological
classification depends on whether the reflection operation ${\cal R}$ commutes
or anticommutes with the non-spatial symmetry operations ${\cal T}$, ${\cal
P}$, and ${\cal C}$. (The condition $U_{\cal R}^2 = 1$ ensures that the 
reflection operation ${\cal R}$ has well-defined algebraic relations with 
${\cal T}$, ${\cal P}$, and ${\cal C}$.)
Following Ref.\ \onlinecite{PhysRevB.88.075142}, to
distinguish the various cases, we use the symbol ${\cal R}$ to denote the
presence of reflection symmetry in the absence of any spatial symmetries,
${\cal R}_{\sigma_{\cal T}}$, ${\cal R}_{\sigma_{\cal P}}$, or ${\cal
R}_{\sigma_{\cal C}}$ to denote a reflection symmetry that commutes (``$\sigma
= +$'') or anticommutes (``$\sigma = -$'') with the non-spatial symmetry
operation ${\cal T}$, ${\cal P}$, or ${\cal C}$ if only one non-spatial
symmetry is present present, and ${\cal R}_{\sigma_{{\cal T}} \sigma_{{\cal
P}}}$ for a reflection symmetry that commutes/anticommutes with time-reversal
symmetry and particle-hole symmetry if all three non-spatial symmetries are
present. (If ${\cal R}$ neither commutes nor anticommutes with the fundamental
non-spatial symmetries, the Hamiltonian $H_d(\vk)$ can be brought into
block-diagonal form, such that there are well-defined commutation or
anticommutation relations between ${\cal R}$ and ${\cal T}$, ${\cal P}$, and
${\cal C}$ for each of the blocks.) The commutation or anticommutation relations between ${\cal R}$ and ${\cal T}$, ${\cal P}$, or ${\cal C}$ imply the algebraic relations $U_{{\cal R}} U_{{\cal T}} = \sigma_{\cal T} U_{{\cal T}} U_{{\cal R}}^*$, $U_{{\cal R}} U_{{\cal P}} = \sigma_{\cal P} U_{{\cal T}} U_{{\cal R}}^*$, and $U_{{\cal R}} U_{{\cal C}} = \sigma_{\cal C} U_{{\cal T}} U_{{\cal C}}$ between the unitary matrices implementing these operations.

\section{Dimensional reduction}\label{sec:methods}

We now describe how one can construct a dimensional
reduction scheme consistent with Bott periodicity using reflection matrices. We
first review how this method works in the absence of reflection symmetry, as
discussed by Fulga {\em et al.},\cite{PhysRevB.85.165409} and then show how to
include the presence of reflection symmetry. 

\subsection{Reflection matrix-based method}

The key step in the method of Ref.~\onlinecite{PhysRevB.85.165409} is the
construction a $(d-1)$-dimensional gapped Hamiltonian $H_{d-1}$ for each
$d$ dimensional gapped Hamiltonian $H_d$. The Hamiltonians $H_d$ and
$H_{d-1}$ have different symmetries, but the same (strong) topological
invariants. Fulga {\em et al.} show how the Hamiltonian $H_{d-1}$ can be
constructed from the reflection matrix $r_d$ if a gapped system with Hamitonian
$H_{d}$ is attached to an ideal lead with a $(d-1)$-dimensional cross section.

Since the reflection matrix depends on the properties of the surface of the
$d$-dimensional insulator, this dimensional reduction method assumes that
the boundary properties can be used to classify the bulk, {\em i.e.},
it assumes a bulk-boundary correspondence. This is the case for the standard
Altland-Zirnbauer classes (without additional symmetries). It is also the case

in the presence of a mirror symmetry, provided the surface contains the normal
to the mirror plane.\cite{PhysRevB.88.075142}

\begin{figure}
\includegraphics[width=0.8\columnwidth]{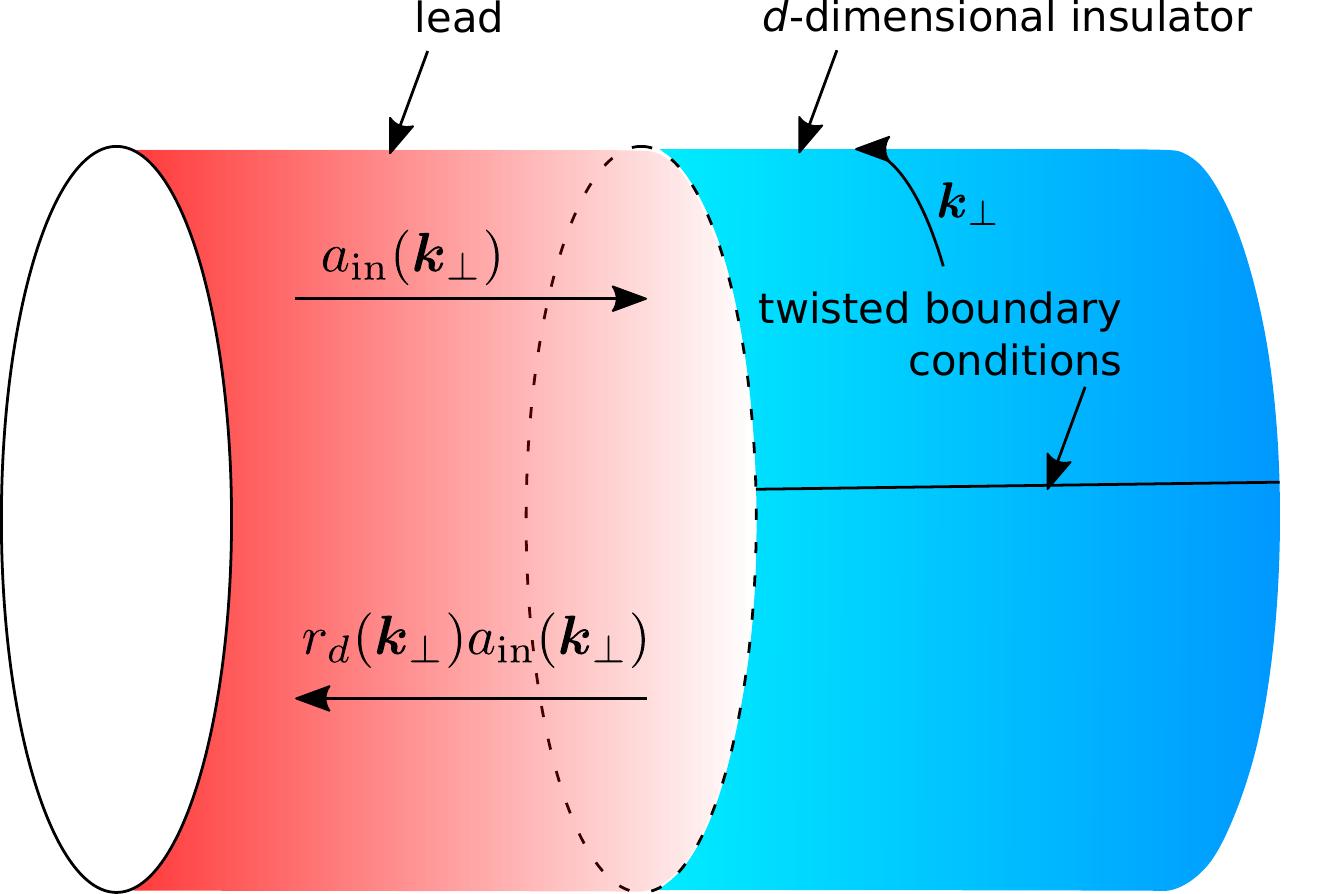}
\caption{\label{fig:setup} Schematic picture of a $d$-dimensional gapped
insulator occupying the half space (blue) with twisted boundary conditions
applied along $(d-1)$-dimension (black line), coupled to an ideal lead (red)
with a $(d-1)$-dimensional cross section. The reflection matrix
$r_d(\vk_{\perp})$ relates the amplitudes of outgoing and incoming modes in the
lead.}
\end{figure}

To be specific, following Ref.\ \onlinecite{PhysRevB.85.165409} we consider a
$d$-dimensional gapped insulator with Hamiltonian $H_d(\vk) =
H_d(k_1,\vk_{\perp})$, occupying the half space $x_1 > 0$, see
Fig.\ \ref{fig:setup}. The half space $x_1
< 0$ consists of an ideal lead with transverse modes labeled by the $d-1$
dimensional wavevector $\vk_{\perp}$. The amplitudes $a_{\rm out}(\vk_{\perp})$
and $a_{\rm in}(\vk_{\perp})$ of outgoing and incoming modes are related by the
reflection matrix $r_d(\vk_{\perp})$,
\begin{equation}
  a_{\rm out}(\vk_{\perp}) = r_d(\vk_{\perp}) a_{\rm in}(\vk_{\perp}).
\end{equation}
Since $H_d$ is gapped, $r(\vk_{\perp})$ is unitary. Time-reversal symmetry,
particle-hole symmetry, or chiral symmetry pose additional constraints on
$r_d(\vk_{\perp})$. These follow from the action of these symmetries on the
amplitudes $a_{\rm in}$ and $a_{\rm out}$,
\begin{align}
  {\cal T} a_{\rm in}(\vk_{\perp}) & = Q_{{\cal T}}\, a_{\rm out}^*(-\vk_{\perp}), \nonumber \\   {\cal T} a_{\rm out}(\vk_{\perp}) &=\, V_{{\cal T}} a_{\rm in}^*(-\vk_{\perp}), \\
  {\cal P} a_{\rm in}(\vk_{\perp}) & = V_{{\cal P}}\, a_{\rm in}^*(-\vk_{\perp}), \nonumber \\  {\cal P} a_{\rm out}(\vk_{\perp}) &= Q_{{\cal P}}\, a_{\rm out}^*(-\vk_{\perp}), \\
  {\cal C} a_{\rm in}(\vk_{\perp}) & = Q_{{\cal C}}\, a_{\rm out}(\vk_{\perp}), \nonumber \\   {\cal C} a_{\rm out}(\vk_{\perp}) & = V_{{\cal C}}\, a_{\rm in}(\vk_{\perp}),
\end{align}
where $V_{{\cal T}}$, $Q_{{\cal T}}$, $V_{{\cal P}}$, $Q_{{\cal P}}$, $V_{{\cal
C}}$, $Q_{{\cal C}}$ are $\vk_{\perp}$-independent unitary matrices that satisfy $V_{\cal T} Q_{{\cal T}}^* = {\cal
T}^2$,  $V_{{\cal P}} V_{{\cal P}}^* = Q_{{\cal P}} Q_{{\cal P}}^* = {\cal
P}^2$, and $Q_{{\cal C}} V_{{\cal C}} = {\cal C}^2 = 1$. Systems with both
time-reversal and particle-hole symmetry also have a chiral symmetry, with
$Q_{{\cal C}} = V_{{\cal P}} Q_{{\cal T}}^* = {\cal T}^2 {\cal P}^2 Q_{{\cal
T}} Q_{{\cal P}}^*$ and $V_{{\cal C}} = Q_{{\cal P}} V_{{\cal T}}^*= {\cal T}^2
{\cal P}^2 V_{{\cal T}} V_{{\cal P}}^*$. For the reflection matrix
$r_d(\vk_{\perp})$ one then finds that the presence of time-reversal symmetry,
particle-hole symmetry, and/or chiral symmetry leads to the constraints
\begin{align}
  r(\vk_{\perp}) = Q_{{\cal T}}^{\rm T} r(-\vk_{\perp})^{\rm T} V_{{\cal T}}^*,
  \label{eq:rtrs} \\
  r(\vk_{\perp}) = Q_{{\cal P}}^{\rm T} r(-\vk_{\perp})^{*} V_{{\cal P}}^*,
  \label{eq:rphs} \\
  r(\vk_{\perp}) = Q_{{\cal C}}^{\dagger} r(\vk_{\perp})^{\dagger} V_{{\cal C}},
  \label{eq:rchiral}
\end{align}

The effective Hamiltonian $H_{d-1}$ is constructed out of $r(\vk_{\perp})$ in
different ways, depending on the presence or absence of chiral symmetry.
With chiral symmetry one sets
\begin{equation}
  H_{d-1}(\vk) \equiv Q_{{\cal C}} r(\vk),
  \label{eq:Hdchiral}
\end{equation}
using Eq.\ (\ref{eq:rchiral}) to verify that $H_{d-1}$ is indeed hermitian.
(Recall that $V_{{\cal C}} = Q_{{\cal C}}^{\dagger}$ since $Q_{{\cal C}} V_{{\cal
C}} = {\cal C}^2 = 1$.) 
Without chiral symmetry one defines $H_{d-1}$ as
\begin{equation}
  H_{d-1}(\vk) = 
  \begin{pmatrix}
    0 & r(\vk)\\
    r^\dagger(\vk) & 0
  \end{pmatrix},
\end{equation}
which is manifestly hermitian and satisfies a chiral symmetry with $U_{\cal C}
= \sigma_3$.

Bulk-boundary correspondence implies that the bulk, which is described by the
Hamiltonian $H_d(\vk)$, and the boundary, which determines the reflection
matrix $r_d(\vk_{\perp})$, have the same topological classification. Since
$r_d(\vk_{\perp})$ is one-to-one correspondence with the Hamiltonian
$H_{d-1}(\vk_{\perp})$, this implies that $H_d$ and $H_{d-1}$ have the same
topological classification (provided bulk-boundary correspondence applies).
Inspecting the symmetries of the sequence of Hamiltonians that results upon
stepwise lowering the dimension $d$, one recovers two periodic sequences of
Hamiltonians with the same topological classification. The appearance of a
period-two sequence for the complex classes
$$
  \mbox{A} \xrightarrow{d-1} \mbox{AIII} \xrightarrow{d-1} \mbox{A}
$$%
follows immediately from the alternating presence and absence of chiral
symmetry in the sequence of Hamiltonians $H_d$ constructed above. To establish
the period-eight sequence one needs to inspect how the dimensional reduction
affects the symmetries of $H$ if time-reversal symmetry and/or particle-hole
symmetry are present. If $H_d$ has both time-reversal and
particle-hole symmetry, $H_{d-1}$ is given by Eq.\ (\ref{eq:Hdchiral}). From
Eqs.\ (\ref{eq:rtrs}), (\ref{eq:rphs}), and (\ref{eq:Hdchiral}) one derives
that time-reversal symmetry and particle-hole symmetry of the reflection matrix
$r_d(\vk)$ yield identical symmetry constraints for the Hamiltonian
$H_{d-1}(\vk)$, 
\begin{equation}
  H_{d-1}(\vk) = {\cal T}^2 {\cal P}^2 V_{{\cal P}}^{\rm T} H_{d-1}^*(-\vk) V_{{\cal P}}^*,
  \label{eq:Hd1TP}
\end{equation}
which has the form of a particle-hole symmetry if ${\cal T}^2 {\cal P}^2 = -
1$, and of a time-reversal symmetry otherwise. In both cases the symmetry
operation squares to $V_{{\cal P}} V_{{\cal P}}^* = {\cal P}^2$. If $H_d$ has
time-reversal symmetry but no particle-hole symmetry, one verifies that
$H_{d-1}$ satisfies
\begin{align}
  H_{d-1}(\vk) & = 
  U^{\dagger} H_{d-1}(-\vk)^*  U,
  \nonumber \\ &=
  -(\sigma_3 U)^{\dagger} H_{d-1}(-\vk)^* (\sigma_3 U),
\end{align}
with
\begin{equation}
  U = \begin{pmatrix}
    0 & V_{{\cal T}}^{*} \\
    Q_{{\cal T}}^{*} & 0
  \end{pmatrix}
\end{equation}
which has the form of a time-reversal symmetry squaring to ${\cal T}^2$ and a
particle-hole symmetry squaring to $-{\cal T}^2$, whereas if $H_d$ has
particle-hole symmetry but no time-reversal symmetry, $H_{d-1}$ satisfies the
symmetry constraints
\begin{align}
  H_{d-1}(\vk) & = 
  {\cal P}^2
  U^{\dagger} H_{d-1}(-\vk)^*  U,
  \nonumber \\ &=
  - {\cal P}^2
  (\sigma_3 U)^{\dagger} H_{d-1}(-\vk)^* (\sigma_3 U),
\end{align}  
with
\begin{equation}
  U = \begin{pmatrix}
    Q_{{\cal P}}^{*} & 0 \\
    0 & V_{{\cal P}}^{*} 
  \end{pmatrix},
\end{equation}
which has the form of a time-reversal symmetry squaring to ${\cal P}^2$ and a
particle-hole symmetry squaring to ${\cal P}^2$. Combining everything, one
arrives at the sequence of symmetry classes
\begin{align}
  \mbox{CI} &\xrightarrow{d-1}
  \mbox{C} \xrightarrow{d-1}
  \mbox{CII} \xrightarrow{d-1}
  \mbox{AII} \xrightarrow{d-1}
  \mbox{DIII}  \nonumber\\ & \xrightarrow{d-1}
  \mbox{D} \xrightarrow{d-1}
  \mbox{BDI} \xrightarrow{d-1}
  \mbox{AI} \xrightarrow{d-1}
  \mbox{CI},
  \label{eq:8seq_standard}
\end{align}
which is the well-known period-eight Bott periodicity known from the
classification of topological insulators and superconductors.\cite{PhysRevB.78.195125,doi:10.1063/1.3149495,1751-8121-44-4-045001,PhysRevB.85.085103,doi:10.1142/S0219887812500235,kennedy2016}

\subsection{reflection symmetry}

The dimensional reduction based on the calculation of reflection matrices can
also be applied in the presence of a reflection symmetry. As in Sec.\
\ref{sec:reflection}, we take the reflection plane to be perpendicular to the
$x_d$ axis, so that the reflection operator ${\cal R}$ maps the lead-system
interface onto itself. As with the non-spatial symmetries, the action of the
reflection operation on the amplitudes $a_{\rm in}$ and $a_{\rm out}$ of
incoming and outgoing states in the leads involves multiplication with
$\vk_{\perp}$-independent unitary matrices,
\begin{align}
  {\cal R} a_{\rm in}(\vk_{\perp}) & = V_{{\cal R}}\, a_{\rm in}(R \vk_{\perp}), \nonumber \\   {\cal R} a_{\rm out}(\vk_{\perp}) &=\, Q_{{\cal R}} a_{\rm out}(R \vk_{\perp}),
\end{align}
where $R \vk_{\perp} = (k_2,\ldots,k_{d-1},-k_{d})$ denotes the reflected mode
vector. The matrices $V_{{\cal R}}$ and $Q_{{\cal R}}$ satisfy $V_{{\cal R}}
Q_{{\cal R}} = {\cal R}^2 = 1$. The presence of reflection symmetry leads to a
constraint on the reflection matrix,
\begin{equation}
  r(\vk_{\perp}) = Q_{{\cal R}}^{\dagger} r(R \vk_{\perp}) V_{{\cal R}}.
\end{equation}
The algebraic relations involving the matrices $Q_{{\cal R}}$, $V_{{\cal R}}$
depend on whether the reflection operation ${\cal R}$ commutes or anticommutes
with the non-spatial symmetry operations ${\cal T}$, ${\cal P}$, and ${\cal
C}$,
$Q_{{\cal T}} Q_{{\cal R}}^* = \sigma_{{\cal T}} V_{{\cal R}} Q_{{\cal T}}$,
$V_{{\cal T}} V_{{\cal R}}^* = \sigma_{{\cal T}} Q_{{\cal R}} V_{{\cal T}}$,
$V_{{\cal P}} V_{{\cal R}}^* = \sigma_{{\cal P}} V_{{\cal R}} V_{{\cal P}}$,
$Q_{{\cal P}} Q_{{\cal R}}^* = \sigma_{{\cal P}} Q_{{\cal R}} Q_{{\cal P}}$,
$Q_{{\cal C}} Q_{{\cal R}} = \sigma_{{\cal C}} V_{{\cal R}} Q_{{\cal C}}$, and
$V_{{\cal C}} V_{{\cal R}} = \sigma_{{\cal C}} Q_{{\cal R}} V_{{\cal C}}$.

To see how the presence of reflection symmetry affects the dimensional
reduction we first consider the complex classes A and AIII. Starting from a
Hamiltonian $H_d$ in symmetry class A with reflection symmetry ${\cal R}$ we
construct a Hamiltonian $H_{d-1}$ in class AIII as described above and find
that reflection symmetry imposes the constraint
\begin{equation}
  H_{d-1}(\vk_{\perp}) = U_{\cal R}^{\dagger} H_{d-1}(R \vk_{\perp}) U_{\cal R},
\end{equation}
on $H_{d-1}$, with
\begin{equation}
  U_{\cal R} = 
  \begin{pmatrix}
    Q_{{\cal R}} & 0 \\
    0 & V_{{\cal R}} 
  \end{pmatrix}.
\end{equation}
Since $U_{\cal R}$ commutes with $\sigma_3$, we conclude that dimensional
reduction maps the class A$^{{\cal R}}$ to AIII$^{{\cal R}_+}$. A similar
procedure can be applied to a Hamiltonian $H_d$ in class AIII with reflection
symmetry ${\cal R}_{\sigma_{\cal C}}$, with $\sigma_{{\cal C}} = \pm$. In this
case, one finds that dimensional reduction leads to a Hamiltonian $H_{d-1}$ in
class A with the additional constraint
\begin{equation}
  H_{d-1}(\vk_{\perp}) = \sigma_{{\cal C}}
  V_{\cal R}^{\dagger} H_{d-1}(R \vk_{\perp}) V_{\cal R}.
  \label{eq:HAIIIR}
\end{equation}
This constraint has the form of a reflection symmetry if $\sigma_{{\cal C}} =
1$, {\em i.e.}, if ${\cal R}$ commutes with ${\cal C}$, but not if
$\sigma_{{\cal C}} = -1$, {\em i.e.}, if ${\cal R}$ anticommutes with ${\cal
C}$. Instead, if $\sigma_{{\cal C}} = -1$ the constraint (\ref{eq:HAIIIR})
represents the product of a reflection symmetry and a chiral symmetry. We
denote such a combined symmetry operation with the symbol ``${\cal CR}$''. To
complete the analysis, we consider a Hamiltonian $H_d$ in class A with the
${\cal CR}$ symmetry constraint, 
\begin{equation}
  H_d(\vk) = - U_{{\cal CR}}^{\dagger} H_d(R \vk) U_{{\cal CR}},
  \label{eq:HdRC}
\end{equation}
where $R \vk = (k_1,k_2,\ldots,k_{d-1},-k_d)$. On the level of the reflection matrix
$r_d(\vk_{\perp})$ the ${\cal CR}$ symmetry is implemented as
\begin{equation}
  r_d(\vk_{\perp}) = Q_{{\cal CR}}^{\dagger} r_d(R \vk_{\perp})^{\dagger} V_{{\cal CR}},
  \label{eq:rdRC}
\end{equation}
where $Q_{{\cal CR}} V_{{\cal CR}} = 1$ and $R\vk_{\perp} =
(k_2,\ldots,k_{d-1},-k_d)$. Performing the dimensional reduction
scheme to this Hamiltonian $H_d$, one immediately finds that $H_{d-1}$
satisfies the constraint
\begin{equation}
  H_{d-1}(\vk_{\perp}) = U_{\cal R}^{\dagger} H_{d-1}(R \vk_{\perp}) U_{\cal R},
  \label{eq:HARC}
\end{equation}
with
\begin{equation}
  U_{\cal R} = \begin{pmatrix}
    0 & Q_{{\cal CR}} \\
    V_{{\cal CR}} & 0
  \end{pmatrix}.
\end{equation}
Since $U_{\cal R}$ anticommutes with $\sigma_3$, the constraint Eq.\
(\ref{eq:HARC}) has the form of a reflection symmetry that anticommutes with
the chiral symmetry ${\cal C}$. Combining everything, we conclude that, once
the symmetry operation ${\cal CR}$ is added, the dimensional reduction scheme
for the complex classes with reflection symmetry leads to two period-two
sequences,
\begin{align}
  \mbox{A}^\mathcal{R}&\xrightarrow{d-1}
  \mbox{AIII}^{\mathcal{R}_{+}}\xrightarrow{d-1}
  \mbox{A}^\mathcal{R} \label{eq:refl_seq2} \\
  \mbox{A}^\mathcal{CR}&\xrightarrow{d-1}
  \mbox{AIII}^{\mathcal{R}_{-}}\xrightarrow{d-1}
  \mbox{A}^\mathcal{CR}.
\end{align}

The symmetry operation ${\cal CR}$ naturally appears in the dimensional
reduction scheme for the real classes as well. 
As with the standard reflection symmetry we have to distinguish
between the cases ${\cal CR}_{\sigma_{\cal T,\cal P}}$ that the ${\cal CR}$
symmetry operation commutes ($\sigma = +$'') or anticommutes (``$\sigma = -$'')
with the time-reversal or particle-hole symmetry operations, if one of these
symmetries is present. (If both symmetries are present, there is no need to
consider ${\cal CR}$ as a separate symmetry operation.) The relations
(\ref{eq:HdRC}) and (\ref{eq:rdRC}) also apply to the general case. 
If chiral symmetry is present, one has $V_{{\cal CR}} = V_{{\cal C}} V_{{\cal
R}}$ and $Q_{{\cal CR}} = Q_{{\cal C}} Q_{{\cal R}}$. One further has the
algebraic relations $Q_{{\cal T}} Q_{{\cal CR}}^* = \sigma_{{\cal T}} V_{{\cal
CR}} Q_{{\cal T}}$, $V_{{\cal T}} V_{{\cal CR}}^* = \sigma_{{\cal T}} Q_{{\cal
CR}} V_{{\cal T}}$, $V_{{\cal P}} V_{{\cal CR}}^* = \sigma_{{\cal P}} V_{{\cal
CR}} V_{{\cal P}}$, $Q_{{\cal P}} Q_{{\cal CR}}^* = \sigma_{{\cal P}} Q_{{\cal
CR}} Q_{{\cal P}}$. Proceeding as above, one verifies that the dimensional
reduction scheme then leads to four period-eight sequences,
\begin{align}
  \mbox{CI}^{\mathcal{R}_{++}} &\xrightarrow{d-1}
  \mbox{C}^{\mathcal{R}_{+}} \xrightarrow{d-1}
  \mbox{CII}^{\mathcal{R}_{++}} \xrightarrow{d-1}
  \mbox{AII}^{\mathcal{R}_{+}} \nonumber\\
  &\xrightarrow{d-1}
  \mbox{DIII}^{\mathcal{R}_{++}} \xrightarrow{d-1}
  \mbox{D}^{\mathcal{R}_{+}} \xrightarrow{d-1}
  \mbox{BDI}^{\mathcal{R}_{++}} \nonumber\\
  &\xrightarrow{d-1}
  \mbox{AI}^{\mathcal{R}_{+}} \xrightarrow{d-1}
  \mbox{CI}^{\mathcal{R}_{++}},
  \label{eq:refl_seq8} \\
  \mbox{CI}^{\mathcal{R}_{--}} &\xrightarrow{d-1}
  \mbox{C}^{\mathcal{R}_{-}} \xrightarrow{d-1}
  \mbox{CII}^{\mathcal{R}_{--}} \xrightarrow{d-1}
  \mbox{AII}^{\mathcal{R}_{-}} \nonumber\\
  &\xrightarrow{d-1}
  \mbox{DIII}^{\mathcal{R}_{--}} \xrightarrow{d-1}
  \mbox{D}^{\mathcal{R}_{-}} \xrightarrow{d-1}
  \mbox{BDI}^{\mathcal{R}_{--}} \nonumber\\
  &\xrightarrow{d-1}
  \mbox{AI}^{\mathcal{R}_{-}} \xrightarrow{d-1}
  \mbox{CI}^{\mathcal{R}_{--}}, \\
  \mbox{CI}^{\mathcal{R}_{-+}} &\xrightarrow{d-1}
  \mbox{C}^{\mathcal{CR}_{+}} \xrightarrow{d-1}
  \mbox{CII}^{\mathcal{R}_{+-}} \xrightarrow{d-1}
  \mbox{AII}^{\mathcal{CR}_{-}} \nonumber\\
  &\xrightarrow{d-1}
  \mbox{DIII}^{\mathcal{R}_{-+}} \xrightarrow{d-1}
  \mbox{D}^{\mathcal{CR}_{+}} \xrightarrow{d-1}
  \mbox{BDI}^{\mathcal{R}_{+-}} \nonumber\\
  &\xrightarrow{d-1}
  \mbox{AI}^{\mathcal{CR}_{-}} \xrightarrow{d-1}
  \mbox{CI}^{\mathcal{R}_{-+}}, \\
  \mbox{CI}^{\mathcal{R}_{+-}} &\xrightarrow{d-1}
  \mbox{C}^{\mathcal{CR}_{-}} \xrightarrow{d-1}
  \mbox{CII}^{\mathcal{R}_{-+}} \xrightarrow{d-1}
  \mbox{AII}^{\mathcal{CR}_{+}} \nonumber\\
  &\xrightarrow{d-1}
  \mbox{DIII}^{\mathcal{R}_{+-}} \xrightarrow{d-1}
  \mbox{D}^{\mathcal{CR}_{-}} \xrightarrow{d-1}
  \mbox{BDI}^{\mathcal{R}_{-+}} \nonumber\\
  &\xrightarrow{d-1}
  \mbox{AI}^{\mathcal{CR}_{+}} \xrightarrow{d-1}
  \mbox{CI}^{\mathcal{R}_{+-}}.
\end{align}
Details of the derivation are given in Appendix~\ref{sec:DimRedR}. The above
sequences were first derifed by Morimoto and Furusaki but ``skipping'' the classes containing $\mathcal{CR}$ symmetry.\cite{PhysRevB.88.125129} Shiozaki and Sato obtained the relations between K groups that give all the sequences derived here as a special case.\cite{PhysRevB.90.165114}

\section{Topological classification with reflection symmetry}\label{sec:exactseq}

Having established the dimensional reduction scheme, it is sufficient to
consider the case $d=1$ in order to completely classify gapped Hamiltonians
with reflection symmetry. (The dimensional reduction scheme can not be used
down to $d=0$ because there can be no reflection-invariant lead-system
interface in one dimension.) Various methods have been used in the literature
to accomplish this task,\cite{PhysRevB.88.075142,PhysRevB.88.125129,PhysRevB.90.165114} as discussed in the introduction or in the review article Ref.\ \onlinecite{RevModPhys.88.035005}.

To make this paper self-contained, we here include a systematic classification
of reflection-symmetric gapped Hamiltonians for $d=1$. We have chosen to use a
different method than used in
Refs.~\onlinecite{PhysRevB.88.075142,PhysRevB.88.125129,PhysRevB.90.165114},
which makes use of concepts from algebraic topology, using relative homotopy
groups and exact sequences. This method was used by Turner {\em et al.} for
their classification of topological insulators with inversion
symmetry.\cite{PhysRevB.85.165120} In App.~\ref{app:K} we discuss how this
classification method can be directly applied to reflection-symmetric
Hamiltonians in dimensions $d > 1$, without the use of the reflection
matrix-based dimensional reduction scheme (and, hence, without the implicit assumption of bulk-boundary correspondence).

The construction of a topological classification for the Hamiltonians $H_d$
requires a mathematical formalism that endows the space of Hamiltonians with a
group structure. The theory of vector bundles and the ``Grothendieck group'' 
provides such a formal structure,
essentially using the diagonal addition of Hamiltonians as the group addition 
operation. Both concepts are reviewed in a language
accessible to physicists, {\em e.g.}, in Ref.~\onlinecite{Budich} and in 
the appendix of Ref.~\onlinecite{PhysRevB.85.165120}. We here employ 
a more informal language, noting that a formally correct formulation requires 
an interpretation of our statements in the framework of the vector bundles 
and the Grothendieck group. As in the previous Section we use the 
Cartan labels to denote the space of hermitian matrices $H$ with a gapped spectrum for the two complex and eight real Altland-Zirnbauer symmetry classes, see Table \ref{tab:classes}.

\begin{figure}
\includegraphics[width=0.8\columnwidth]{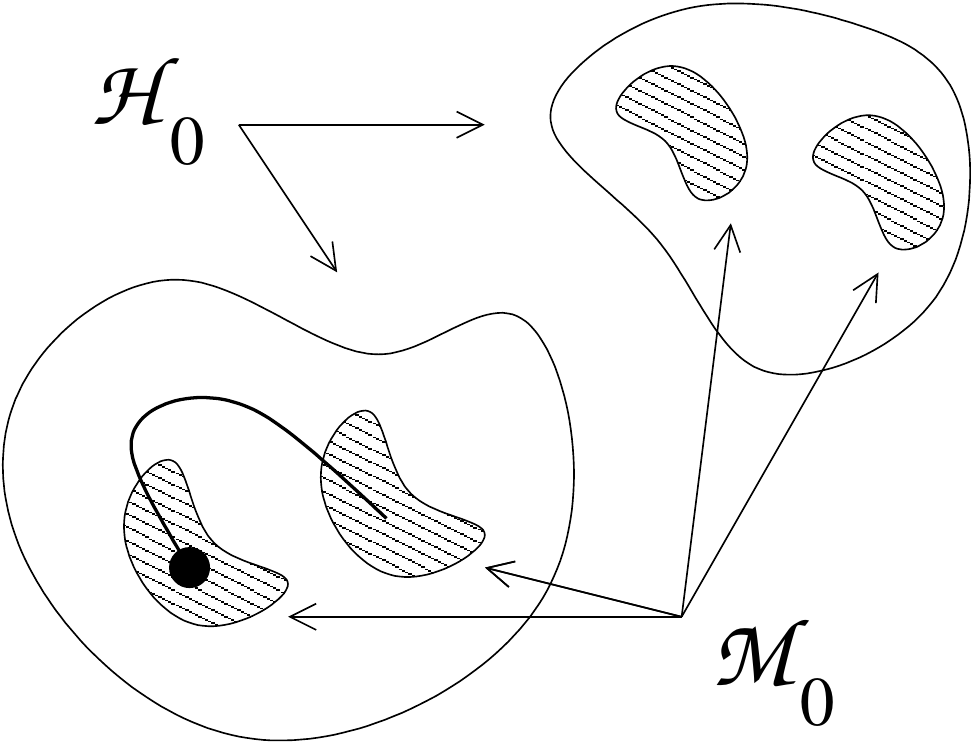}
\caption{\label{fig:illustration} Schematic illustration of the spaces $\cH{0}$
and $\cR{0}$. The solid dot indicates the trivial element. The thick curve
shows a path in $\cH{0}$ that starts at the trivial element and ends in
$\cR{0}$. Equivalence classes of such paths form the relative homotopy group
$\pi_1(\cH{0},\cR{0})$.}
\end{figure}

In one dimension, we are interested in in periodic, functions $H(k)=H(k+2\pi)$,
with $H(k)$ a gapped Hamiltonian, where the antiunitary symmetry operations
${\cal T}$ and ${\cal P}$ as well as the reflection operations ${\cal R}$ and
${\cal CR}$ relate $H(k)$ and $H(-k)$. 
It is then sufficient to consider the Hamiltonian $H(k)$ on the interval $0 \le
k \le \pi$ only. For generic $0 < k < \pi$ only symmetries that relate $H(k)$
to itself play a role. These symmetries confine $H(k)$ for $0 < k < \pi$ to one
of the classifying spaces of table \ref{tab:classes}. We use the symbol
$\cH{0}$ to denote this space. The momenta $k=0$ and $k=\pi$ are mapped to
themselves under $k \to -k$, so that $H(0)$ and $H(\pi)$ satisfy additional
symmetries. We use $\cR{0}$ to denote the classifying space of Hamiltonians
that also satisfy these additional symmetry constraints. Figure
\ref{fig:illustration} schematically illustrates the spaces $\cH{0}$ and
$\cR{0}$.

In general a Hamiltonian $H(k)$ can be block-decomposed as $H(k)=H(0)\oplus
H^\prime(k)$, where $H^\prime(0)$ is topologically ``trivial''. The
$k$-independent Hamiltonian $H(0)$ has topological indices characteristic of
the zero dimensional case. These indices become weak indices of one-dimensional
Hamiltonian $H(k)$. The classification of the Hamiltonians $H'(k)$ gives the
strong topological indices. 

In view these considerations, our goal is to classify functions $H'(k)$ on the
interval $0 \le k \le \pi$, such that $H'(k)$ is gapped, $H'(0)$ is trivial,
$H'(\pi) \in \cR{0}$, and $H(k) \in \cH{0}$ otherwise. The space of equivalence
classes (defined with respect to continuous deformations) of such functions
$H'(k)$ is known as the \textit{relative homotopy group}
$\pi_1(\cH{0},\cR{0})$.~\cite{nash1988topology,PhysRevB.85.165120} The group
$\pi_1(\cH{0},\cR{0})$ gives the topological classification of gapped
Hamiltonians with the desired symmetries. A function $H'(k)$ with these
constraints can be interpreted as a continuous ``path'' in $\cH{0}$, starting
at the trivial point, and ending somewhere in $\cR{0}$, see
Fig.~\ref{fig:illustration}. 

The relative homotopy group can be calculated from the zeroth and first
homotopy groups of $\cH{0}$ and $\cR{0}$, where we recall that the zeroth
homotopy groups $\pi_0(\mathcal{X})$ labels the connected components of a
topological space $\mathcal{X}$, whereas the first homotopy group
$\pi_1(\mathcal{X})$ contain equivalence classes of ``closed loops'' in
$\mathcal{X}$ that begin and end at the trivial reference point.
This calculation makes use of an ``exact sequence'' of mappings\cite{PhysRevB.85.165120}
\begin{align}
  \pi_1(\cR{0}) &\overset{i_1}{\hookrightarrow}\pi_1(\cH{0})\overset{j_1}{\hookrightarrow}\pi_1(\cH{0},\cR{0})
  \nonumber \\ &\overset{\delta}{\rightarrow}\pi_0(\cR{0})\overset{i_0}{\hookrightarrow}\pi_0(\cH{0}),
  \label{eq:exactseq}
\end{align}
where a sequence of mappings is called ``exact'' if the image of each mapping
is the kernel of the subsequent one. In the sequence (\ref{eq:exactseq}), the
maps $i_1$, $j_1$ and $i_0$ are inclusion maps where the same object is interpreted
as an element of a larger space.
The map $\delta$ is the ``boundary map'', mapping an equivalence class of ``pathes'' $H(k)$ in $\pi_1(\cH{0},\cR{0})$ to the connected component of their endpoint $H(\pi)$ in $\cR{0}$. Since the groups $\pi_1(\cH{0})$ and $\pi_0(\cR{0})$, as well as the image of $i_1$ and the kernel of $i_0$ are known, the relative homotopy group $\pi_1(\cH{0},\cR{0})$ and its structure follow immediately from the exactness of the sequence (\ref{eq:exactseq}).
%
Similarly, generators for $\pi_1(\cH{0},\cR{0})$ can be constructed by application of the inclusion map $j_1$ and a suitable inverse of the boundary map $\delta$.
Table \ref{tab:classes} lists the groups $\pi_0$ and $\pi_1$ for the classifying spaces $\mathcal{C}_n$ and $\mathcal{R}_n$. 

To classify one-dimensional gapped Hamiltonians with reflection symmetry, the spaces $\cH{0}$ and $\cR{0}$ are identified for each symmetry class, see Tables \ref{tab:AZcomplex} and \ref{tab:AZreal}, for the two period-two ``complex'' sequences and for the four period-eight ``real'' sequences, respectively. The relative homotopy group $\pi_1(\cH{0},\cR{0})$, which classifies the gapped Hamiltonians with reflection or $\mathcal{CR}$ symmetry, is then calculated from the exact sequence (\ref{eq:exactseq}). The results of this classification are shown in Tables \ref{tab:AZcomplex} and \ref{tab:AZreal}. In addition to the classification for $d=1$, the table also lists the results for $d=2$, $3$, and $4$, following the Bott clock structure outlined in the previous Section. The assignment of the spaces $\cH{0}$ and $\cR{0}$ for the different symmetry classes and the details on the resolution of the exact sequence in the nontrivial cases is discussed in detail in appendix \ref{app:exactseq}.

\begin{table}
  \begin{tabular}{|c||c|c||c|c|c|c|}
    \hline
    class & $\cH{0}^i$ & $\cR{0}^i$ & $d=1$ & $d=2$ & $d=3$ & $d=4$\\\hline
    A$^{\mathcal{R}}$ & AI & AI$^2$ & $\bZ$ & $0$ & $\bZ$ & $0$\\\hline
    AIII$^{\mathcal{R}_+}$ & AIII & AIII$^2$ & $0$ & $\bZ$ & $0$ & $\bZ$\\
    \hline \hline
    A$^{\mathcal{CR}}$ & A & AIII & 0 & ${\bZ^2}$ & $0$ & ${\bZ^2}$\\\hline
    AIII$^{\mathcal{R}_-}$ & AIII & A  & ${\bZ^2}$ & $0$ & ${\bZ^2}$ & 0\\\hline
  \end{tabular}
\caption{  \label{tab:AZcomplex}
The complete classification for the complex Altland-Zirnbauer classes with reflection
symmetry.}
\end{table}

\begin{table}
  \begin{tabular}{|c||c|c||c|c|c|c|}
    \hline
    class & $\cH{0}^i$ & $\cR{0}^i$ & $d=1$ & $d=2$ & $d=3$ & $d=4$\\\hline
    AI$^{\mathcal{R}_+}$ & AI & AI$^2$ & $\bZ$ & $0$ & $0$ & $0$\\\hline
    BDI$^{\mathcal{R}_{++}}$ & BDI & BDI$^2$ & $\bZ_2$ & $\bZ$ & $0$ & $0$\\\hline
    D$^{\mathcal{R}_{+}}$ & D & D$^2$ & $\bZ_2$ & $\bZ_2$ & $\bZ$ & $0$\\\hline
    DIII$^{\mathcal{R}_{++}}$ & DIII & DIII$^2$ & $0$ & $\bZ_2$ & $\bZ_2$ & $\bZ$\\\hline
    AII$^{\mathcal{R}_{+}}$ & AII & AII$^2$ & $2\bZ$ & $0$ & $\bZ_2$ & $\bZ_2$\\\hline
    CII$^{\mathcal{R}_{++}}$ & CII & CII$^2$ & $0$ & $2\bZ$ & $0$ & $\bZ_2$\\\hline
    C$^{\mathcal{R}_{+}}$ & C & C$^2$ & $0$ & $0$ & $2\bZ$ & $0$\\\hline
    CI$^{\mathcal{R}_{++}}$ & CI & CI$^2$ & $0$ & $0$ & $0$ & $2\bZ$\\
    \hline \hline
    AI$^{\mathcal{R}_-}$ & AII & A & $0$ & $0$ & $2\bZ$ & $0$\\\hline
    BDI$^{\mathcal{R}_{--}}$ & CII & AIII &$0$ & $0$ & $0$ & $2\bZ$\\\hline
    D$^{\mathcal{R}_{-}}$ & C & A & $\bZ$ & $0$ & $0$ & $0$\\\hline
    DIII$^{\mathcal{R}_{--}}$ & CI & AIII & $\bZ_2$ & $\bZ$ & $0$ & $0$\\\hline
    AII$^{\mathcal{R}_{-}}$ & AI & A & ${\bZ_2}$ & $\bZ_2$ & $\bZ$ & $0$\\\hline
    CII$^{\mathcal{R}_{--}}$ & BDI & AIII & $0$ & ${\bZ_2}$ & $\bZ_2$ & $\bZ$\\\hline
    C$^{\mathcal{R}_{-}}$ & D & A & $2\bZ$ & $0$ & ${\bZ_2}$ & $\bZ_2$\\\hline
    CI$^{\mathcal{R}_{--}}$ & DIII & AIII & $0$ & $2\bZ$ & $0$ & ${\bZ_2}$\\
    \hline \hline
    AI$^{\mathcal{CR}_-}$ & C & CI & $0$ & $0$ & $0$ & ${2\bZ^2}$\\\hline
    BDI$^{\mathcal{R}_{+-}}$ & CI & AI & ${\bZ^2}$ & $0$ & $0$ & $0$\\\hline
    D$^{\mathcal{CR}_{+}}$ & AI & BDI & ${\bZ_2^2}$ & ${\bZ^2}$ & $0$ & $0$\\\hline
    DIII$^{\mathcal{R}_{-+}}$ & BDI & D & ${\bZ_2^2}$ & ${\bZ_2^2}$ & ${\bZ^2}$ & $0$\\\hline
    AII$^{\mathcal{CR}_{-}}$ & D & DIII & $0$ & ${\bZ_2^2}$ & ${\bZ_2^2}$ & ${\bZ^2}$\\\hline
    CII$^{\mathcal{R}_{+-}}$ & DIII & AII & ${2\bZ^2}$ & $0$ & ${\bZ_2^2}$ & ${\bZ_2^2}$\\\hline
    C$^{\mathcal{CR}_{+}}$ & AII & CII & $0$ & ${2\bZ^2}$ & $0$ & ${\bZ_2^2}$\\\hline
    CI$^{\mathcal{R}_{-+}}$ & CII & C & $0$ & $0$ & ${2\bZ^2}$ & $0$\\
    \hline \hline
    AI$^{\mathcal{CR}_+}$ & D & BDI & $0$ & $2\bZ$ & $0$ & $\bZ$\\\hline
    BDI$^{\mathcal{R}_{-+}}$ & DIII & D & $\bZ$ & $0$ & $2\bZ$ & $0$\\\hline
    D$^{\mathcal{CR}_{-}}$ & AII & DIII & $0$ & $\bZ$ & $0$ & $2\bZ$\\\hline
    DIII$^{\mathcal{R}_{+-}}$ & CII & AII & $2\bZ$ & $0$ & $\bZ$ & $0$\\\hline
    AII$^{\mathcal{CR}_{+}}$ & C & CII & $0$ & $2\bZ$ & $0$ & $\bZ$\\\hline
    CII$^{\mathcal{R}_{-+}}$ & CI & C & $\bZ$ & $0$ & $2\bZ$ & $0$\\\hline
    C$^{\mathcal{CR}_{-}}$ & AI & CI & $0$ & $\bZ$ & $0$ & $2\bZ$\\\hline
    CI$^{\mathcal{R}_{+-}}$ & BDI & AI & $2\bZ$ & $0$ & $\bZ$ & $0$\\\hline
  \end{tabular}
\caption{The complete classification for the real Altland-Zirnbauer classes with reflection
symmetry.}
  \label{tab:AZreal}
\end{table}

\section{The second descendant $\bZ_2$ phase} \label{sec:Z2}

Chiu {\em et al.}\cite{PhysRevB.88.075142} and Morimoto and
Furusaki\cite{PhysRevB.88.125129} argued that the class CII$^{\mathcal{R}_{--}}$
of reflection-symmetric topological superconductors in two dimensions ($d=2$)
has gapless boundary states that not protected against perturbations that lift
the discrete translation symmetry of the underlying lattice. On the other hand,
Shiozaki and Sato point out that this class has a well-defined strong index,
although they nevertheless allow for a ``subtle instability'' of the
topologically nontrivial state.\cite{PhysRevB.90.165114}

The dimensional reduction scheme links class CII$^{\mathcal{R}_{--}}$ with $d=2$
to class AII$^{\mathcal{R}_-}$ in one dimension, {\em i.e.}, the reflection
matrix $r_2(k_{\perp})$ of a two-dimensional Hamiltonian $H_2(\vk)$ in class
CII$^{\mathcal{R}_{--}}$ is a one-dimensional object with symmetries
characteristic of class AII$^{\mathcal{R}_-}$. In this Section we show that the
definition of the $\bZ_2$ topological invariant for class AII$^{\mathcal{R}_-}$
is robust to the addition of perturbations that break the discrete translation
symmetry, consistent with the observation of Shiozaki and Sato that there is a
well-defined topological index.\cite{PhysRevB.90.165114} We then use our
scattering approach to show that a nontrivial value of the invariant implies
the existence of gapless states at the boundary of the two-dimensional system.
As explained in App.\ \ref{sec:simulations}, we believe the fact that Refs.\
\onlinecite{PhysRevB.88.075142} and \onlinecite{PhysRevB.88.125129} observe a
gap opening for edge states is because the perturbation considered there
includes a long-range hopping term with a hopping amplitude decaying inversely
proportional to distance, which does not result in a continuous Bloch
Hamiltonian $H(\vk)$ as a function of $\vk$.

The class AII$^{\mathcal{R}_-}$ has time-reversal symmetry with
$\mathcal{T}^2=-1$. Combining the reflection and time-reversal symmetries we
arrive at
\begin{align}
  (\mathcal{RT})H(k)(\mathcal{RT})=H(k),
  \label{eq:rt}
\end{align}
with $(\mathcal{RT})^2=1$ since $\mathcal{R}$ and $\mathcal{T}$ anticommute.
Without loss of generality we may represent $\mathcal{RT}$ by complex
conjugation $K$ and $\mathcal{R}$ by $\sigma_2$, so that $H(k)$ is a real
symmetric matrix with the additional condition $H(k) = \sigma_2 H(-k)
\sigma_2$. We conclude that $\cH{0}$ is the class AI, whereas at the reflection
symmetric points $k=0$, $\pi$, $H$ is of the form
\begin{equation} H = \begin{pmatrix} a & b \\ -b & a \end{pmatrix}\,
  \label{eq:complex}
\end{equation}
with $a$ ($b$) real symmetric (antisymmetric). Such a structure forms a
two-dimensional representation of the complex numbers, so that we find that
$\cR{0}$ is the space of gapped Hamiltonians in class A. Following the general
procedure of Sec.\ \ref{sec:exactseq} we write $H(k) = H(0) \oplus H'(k)$,
where $H'(0)$ is topologically trivial. This gives the exact sequence
\begin{align}
  0\overset{i_1}{\hookrightarrow}\bZ_2\overset{j_1}{\hookrightarrow}\bZ_2\overset{\delta}{\rightarrow}\bZ\overset{i_0}{\hookrightarrow}\bZ,
  \label{eq:a2mpexact}
\end{align}
with $\pi_1(\cH{0},\cR{0})=\bZ_2$. The topological structure is inherited from
the left part of the exact sequence. The index can be calculated as the standard
invariant classifying loops of real symmetric matrices,\cite{PhysRevB.76.045302}
\begin{equation}
  W' = \prod_{k=0,\pi} \mbox{Pf}\, w(k),
  \label{eq:A2inv}
\end{equation}
with
\begin{equation}
  w_{ab}(k)= - w_{ba}(k) = 
  - i \langle u(-k)_a\vert\mathcal{T}\vert u_b(k)\rangle,
  \label{eq:w}
\end{equation}
the $\vert u_a(k)\rangle$ being the Bloch wave functions of occupied states.
(Recall that the time-reversal operation ${\cal T}$ maps $k$ to $-k$.) At the
reflection-symmetric points $k=0$, $\pi$, the matrix $w(k)$ is orthogonal
and skew-symmetric, so that the Pfaffian can only take values the $\pm1$. 

The Hamiltonian
\begin{equation}
  H^{(-1)}(k) = \begin{pmatrix} \cos k \tau_3 + \sin k \tau_1 & 0 \\ 0 & \cos k \tau_3 - \sin k \tau_1 \end{pmatrix}
  \label{eq:Ha2}
\end{equation}
is a generator with $W' = -1$. This can be seen, noting that the eigenvectors
of the negative-energy states are
\begin{equation}
  \vert u_1(k)\rangle = \begin{pmatrix}
  \sin(k/2) \\ -\cos(k/2) \\ 0 \\ 0 \end{pmatrix},\ \
  \vert u_2(k)\rangle = \begin{pmatrix} 0 \\ 0 \\ \sin(k/2) \\ \cos(k/2)
  \end{pmatrix},
\end{equation}
from which it follows that 
\begin{equation}
  w(k) = \begin{pmatrix} 0 & \cos k \\ - \cos k & 0 \end{pmatrix}
\end{equation}
and $W' = -1$.

At this point we like to stress again that in the classification procedure weak
topological numbers, {\em i.e.}, topological numbers that are not robust
against translation-symmetry breaking, are associated with $H(0)$, not with
$H'(k)$. Hence, the invariant $W'$ should be a {\em strong} invariant.  To see
this explicitly, we now show that the nontrivial topological index for the
Hamiltonian $H^{(-1)}(k)$ of Eq.~(\ref{eq:Ha2}) survives the redefinition of
the unit cell that comes with the breaking of translation symmetry. Indeed, out
of Eq.~(\ref{eq:Ha2}) one can construct an eight-band Hamiltonian
${H}_2^{(-1)}(k)$ as
\begin{equation}
  {H}_2^{(-1)}(k) = e^{-i \mu_2 k/2} [H^{(-1)}(k) \oplus H^{(-1)}(k+\pi)] e^{i \mu_2 k/2},
\end{equation}
where the Pauli matrix $\mu_2$ acts in the space consisting of the two original
unit cells that form the new unit cell. One verifies that the unit cell for
$H_2^{(-1)}(k)$ is twice that of the original Hamiltonian $H^{(-1)}(k)$, {\em
i.e.}, that ${H}_2^{(-1)}(k) = {H}_2^{(-1)}(k+\pi)$, and that ${H}_2^{(-1)}(k)$
has the same symmetries as $H^{(-1)}(k)$, defined by
$\mathcal{R}=\mu_3\sigma_2$ and $\mathcal{RT}=K$. (The fact that the expression
for the reflection differs from that for the original unit cell is natural,
since a reflection plane that maps the original unit cell to itself no longer
maps a doubled unit cell to itself.) A direct calculation gives that the
invariant (\ref{eq:A2inv}) is the same for both $H^{(-1)}(k)$ and
${H}_2^{(-1)}(k)$. Similarly, one may construct Hamiltonians in which the unit
cell size is multiplied by the odd number, with the same properties see
Appendix~\ref{app:folding}. We conclude that the invariant (\ref{eq:A2inv}) is
a true strong invariant.

\begin{figure}
\includegraphics[width=0.8\columnwidth]{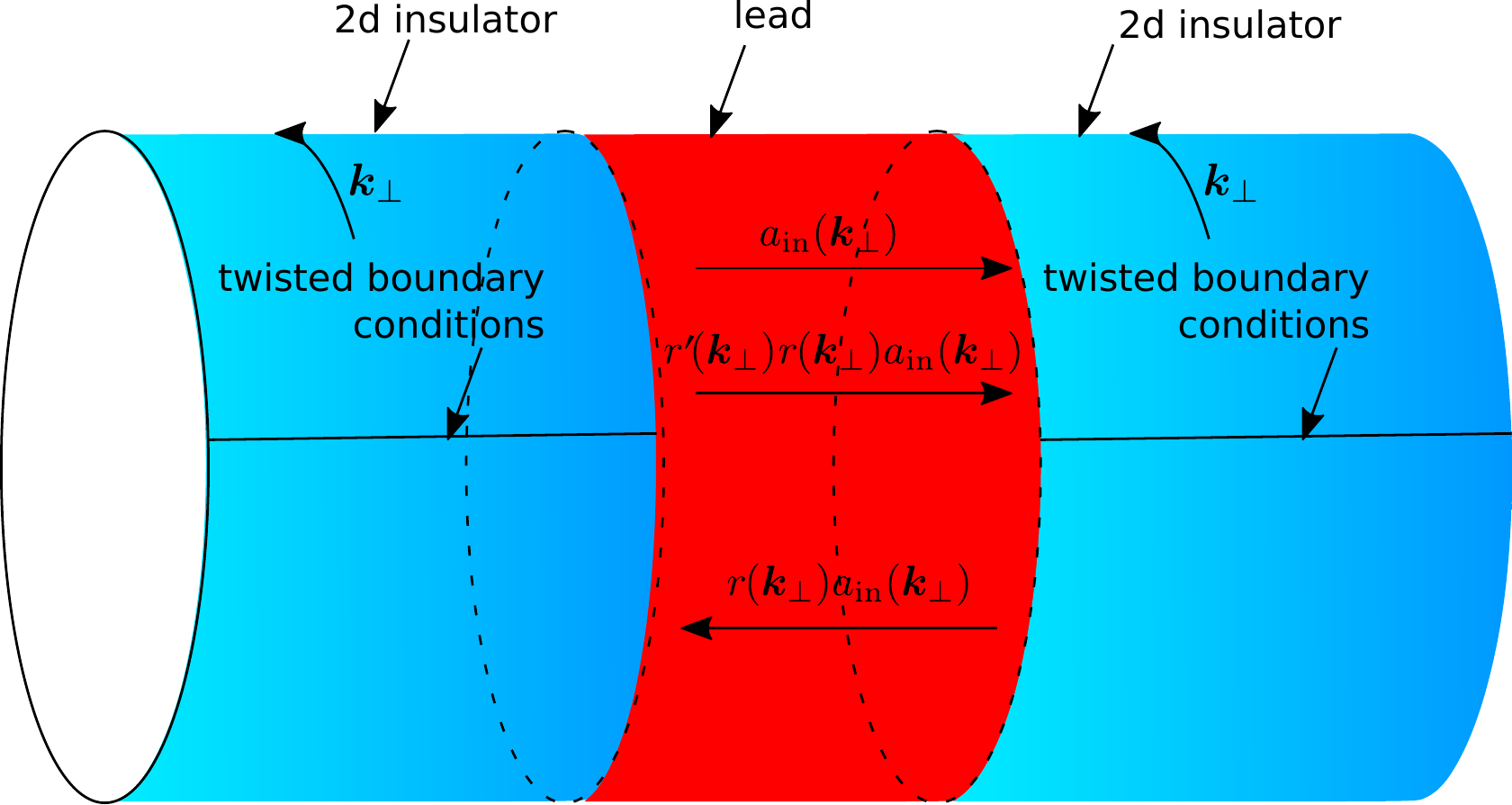}
\caption{\label{fig:CII} Schematic illustration of a two-dimensional insulator,
attached to an ideal lead of finite length. The insulator has reflection matrix
$r$, the lead is terminated with reflection matrix $r'$. This setup is used to
obtain the boundary state from the scattering approach.}
\end{figure}

It remains to show that a nontrivial value of the topological invariant implies
the existence of a gapless boundary state. To relate the reflection matrix to
boundary states, we consider a two-dimensional topological insulator in class
CII$^{\mathcal{R}_{--}}$ and attach an ideal lead to the left, see Fig.\
\ref{fig:CII}. As discusssed previously, the reflection matrix $r(k_{\perp})$
(when multiplied by $Q_{\cal C}$) belongs to symmetry class
AII$^{\mathcal{R}_{-}}$. To model a sample edge, the lead is terminated on its
left end by a reflection matrix $r'(k_{\perp})$ which, again, is in symmetry
class AII$^{\mathcal{R}_{-}}$ (when multiplied by $Q_{\cal C}$). The condition
to have a boundary state at zero energy is
\begin{align}
  \det(\openone - r(k_\perp)r'(k_\perp))=0. \label{eq:boundstates}
\end{align}
If the two-dimensional bulk is in a nontrivial topological class in
CII$^{\mathcal{R}_{--}}$, the two matrices $r(k_\perp)$ and $r'(k_\perp)$ are
in {\em different} topological classes in AII$^{\mathcal{R}_{-}}$. We now show
that this is sufficient to ensure the existence of zero-energy boundary states,
{\em i.e.}, that Eq.\ (\ref{eq:boundstates}) has a solution for at least one
value of $k_{\perp}$.

We first note that $Q_{\cal C} r(k_\perp)$ and $r'(k_\perp) Q_{\cal C}$ are
hermitian and unitary, so that all eigenvalues are $1$ or $-1$, the number of
negative eigenvalues being the same for all $k_{\perp}$. Moreover, at $k=0$ and
$k=\pi$ $Q_{\cal C} r(k_\perp)$ and $r'(k_\perp) Q_{\cal C}$ all eigenvalues
are twofold degenerate because of the block structure (\ref{eq:complex}). If
the number of positive and negative eigenvalues are not balanced, the product
$r(k_\perp)r'(k_\perp)$ trivially has a unit eigenvalue for all $k_{\perp}$,
{\em i.e.}, there is a ``flat band'' of surface states. Hence, we may restrict
ourselves to the case that the number of positive and negative eigenvalues are
equal. This implies that the dimension of $r$ and $r'$ is $4N$, with $N$
integer, as the eigenvalues of $Q_{\cal C} r(k_\perp)$ and $r'(k_\perp) Q_{\cal
C}$ are twofold degenerate at $k_{\perp} = 0$, $\pi$. 

We take real bases $\{ |u_{j}(k_{\perp}) \rangle \}$ and $\{ |u_{j}'(k_{\perp})
\rangle \}$ for the negative-eigenvalue eigenspaces of $Q_{\cal C}r(k_{\perp})$
and $r'(k_{\perp}) Q_{\cal C}$, respectively, with the constraint that
\begin{equation}
  |u_{2l}(0)\rangle = i {\cal T}
  |u_{2l-1}(0)\rangle,\ \ l = 1,2,\ldots,N,
  \label{eq:c2}
\end{equation}
and similar for the primed basis set. This gives $\mbox{Pf}\,(w(0)) =
\mbox{Pf}\,(w'(0)) = 1$, where $w$ and $w'$ are the antisymmetric matrices
defined for $Q_{\cal C}r(k_{\perp})$ and $r'(k_{\perp}) Q_{\cal C}$,
respectively, see Eq.\ (\ref{eq:w}). We then define the $4 N \times 4N$ real
matrix $O(k_{\perp})$ with the first (second) $2 N$ columns given by the
vectors $|u_{j}(k_{\perp}) \rangle$ ($\{ |u_{j}'(k_{\perp}) \rangle$),
$j=1,2,\ldots,2N$. Below we show that either $\det O(0) \det O(\pi) = 0$ or
\begin{equation}
  \mbox{sign}\, \det O(0) \det O(\pi) = W'_{r} W'_{r'},
  \label{eq:OO}
\end{equation}
where $W'_{r}$ and $W'_{r'}$ are the topological numbers for $Q_{\cal
C}r(k_{\perp})$ and $r'(k_{\perp}) Q_{\cal C}$, respectively. In both cases it
follows that $\det O(k_{\perp})$ is zero for at least one value of $k_{\perp}$,
{i.e.}, the vectors $\{ |u_{j}(k_{\perp}) \rangle,\, |u_{j}'(k_{\perp}) \rangle
\}$ are linearly depenent and, hence, do not span the full $4N$-dimensional
space. Since 
\begin{align}
  Q_{\cal C}r(k_{\perp}) &= \openone - 2 \sum_{j=1}^{2N}
  | u_{j}(k_{\perp}) \rangle \langle u_{j}(k_{\perp})|
  \nonumber \\
  r'(k_{\perp}) Q_{\cal C} &= \openone - 2 \sum_{j=1}^{2N}
  | u_{j}'(k_{\perp}) \rangle \langle v_{j}(k_{\perp})|
\end{align}
it follows that $r(k_{\perp}) r'(k_{\perp})$ has an eigenvalue one and, thus,
Eq.\ (\ref{eq:boundstates}) guarantees the existence of a boundary state for
that value of $k_{\perp}$.

Equation (\ref{eq:OO}) follows immediately if the negative-eigenvalue subspaces
of $r'(k_{\perp}) Q_{\cal C}$ and $Q_{\cal C}r(k_{\perp})$ each are the same at
$k_{\perp} = 0$ and $k_{\perp} = \pi$. In that case, the topological numbers
$W_{r}'$ and $W_{r'}'$ give the ``handedness'' of the transformation linking
the basis states at $k_{\perp} = 0$ and at $k_{\perp} = \pi$,
\begin{equation}
  W_{r}' = \mbox{sign}\, \det[\langle u_i(\pi)| u_j(0) \rangle],
\end{equation}
with a similar equality for $W_{r'}'$. The determinant $\det O(\pi)$ gives the
relative handedness of the two transformations, if the basis sets $\{
|u_{j}(k_{\perp}) \rangle \}$ and $\{ |v_{j}(k_{\perp}) \rangle \}$ are
linearly independent at $k_{\perp} = 0$, $\pi$, so that Eq.\ (\ref{eq:OO})
follows. (If the basis sets $\{ |u_{j}(k_{\perp}) \rangle \}$ and $\{
|v_{j}(k_{\perp}) \rangle \}$ are not linearly independent at $k_{\perp} =
0$, $\pi$, one has $\det O(0) \det O(\pi) = 0$.) In the general case, one
can show that Eq.\ (\ref{eq:OO}) holds by comparing the basis sets $\{
|u_{j}(k_{\perp}) \rangle \}$ and $\{ |u_{j}'(k_{\perp}) \rangle \}$ at
$k_{\perp} = \pi$ with two real reference basis sets $\{ |\tilde u_{j}(s)
\rangle \}$ and $\{ |\tilde u_{j}'(s) \rangle \}$, $0 \le s \le \pi$, which
are identical to the original sets $\{ |u_{j}(k_{\perp}) \rangle \}$ and
$\{ |u_{j}'(k_{\perp}) \rangle \}$ at $k_{\perp} = 0$, span the same
subspaces as the original sets $\{ |u_{j}(k_{\perp}) \rangle \}$ and $\{
|u_{j}'(k_{\perp}) \rangle \}$ at $k_{\perp} = \pi$, and satisfy the
constraint (\ref{eq:c2}) at all values of $0 \le s \le \pi$. (Here the
time-reversal operation does not send $s$ to $-s$.) The topological
invariant $W_{r}'$ can then be calculated as the ``handedness'' of the
transformation between the basis sets $\{ |u_{j}(\pi) \rangle \}$ and $\{
|\tilde u_{j}(\pi) \rangle \}$,
\begin{equation}
  W_{r}' = \mbox{sign}\, \det[\langle u_i(\pi)|\tilde u_j(\pi) \rangle],
\end{equation}
with a similar result for the topological invariant $W_{r'}'$. The desired
result now follows from the observation that shifting the reference basis sets
from $s = \pi$ to $s=0$ does not change the handesness of the transformation. In
Appendix~\ref{sec:simulations} we carry out the analysis of the present section,
numerically on a concrete example.

\section{Summary} \label{sec:summary}
We have studied the classification of topological insulators and superconductors
in the presence of reflection symmetry. We used method based on the reflection 
matrix to derive the Bott clock which is in agreement with previous
works.~\cite{PhysRevB.88.125129,PhysRevB.90.165114} For the sake of completeness
we also obtained the classification using method based on relative homotopy
groups and exact sequences; our results are in full agreement with those of
Shizoaki \textit{et. al.}~\cite{PhysRevB.90.165114} and partial agreement with
Refs.~\onlinecite{PhysRevB.88.075142,PhysRevB.88.125129}. We also show that the
non-trivial topological phases classified with second descendant $\bZ_2$ are
robust to disorder.

The dimensional reduction scheme based on the reflection matrix is distinguished
by being physically intuitive since it relates the topological invariant to
transport properties of a system. Additionally it offers a high computational
efficiency when studying the effects of disorder since one can consider systems
in lower dimension. The dimensional reduction method used in this work is purely
algebraic and can be readily extended to other point group symmetries; together
with relative exact sequences this method could yield the complete
classification of topological phases of matter.

\acknowledgements

We thank Akira Furusaki, Cosma Fulga, and Shinsei Ryu for stimulating discussions. We acknowledge support by project A03 of the CRC-TR 183 and by the priority programme SPP 1666 of the German Science Foundation (DFG).

\appendix

\section{Dimensional reduction with reflection symmetry}\label{sec:DimRedR}

Here we give the details of the calculation of four period-eight sequences for
the real classes with reflection symmetry. We need to distinguish three cases
separately: ({\em i}) $H_d$ has both time-reversal and particle-hole symmetry,
({\em ii}) $H_d$ has time-reversal symmetry, but no particle-hole symmetry, and
({\em iii}) $H_d$ has particle-hole symmetry, but no time-reversal symmetry.

({\em i}).---
If $H_d$ has both time-reversal and particle-hole symmetry, then $H_{d-1}$
satisfies
\begin{align}
  H_{d-1}(\vk) &= {\cal T}^2 {\cal P}^2 V_{{\cal P}}^{\rm T} H_{d-1}^*(-\vk)
  V_{{\cal P}}^*,\\
  H_{d-1}(\vk) &=
  \sigma_\mathcal{C}V_\mathcal{R}^\dagger H_{d-1}(R\vk)V_\mathcal{R}.
  \label{eq:PTred}
\end{align}
The first constraint has the form of a particle-hole symmetry if
$\mathcal{T}^2\mathcal{P}^2=-1$, and of a time-reversal symmetry otherwise. The
second constraint has the form of reflection symmetry ($\mathcal{CR}$ symmetry)
if $\sigma_\mathcal{C}=1$ ($\sigma_\mathcal{C}=-1$), with the algebraic relation
to the non-spatial symmetry ($\mathcal{T}$ or $\mathcal{P}$) given by
$\sigma_\mathcal{P}$. 

({\em ii}).---
If $H_d$ has time-reversal symmetry but no particle-hole
symmetry, one has
\begin{align}
  H_{d-1}(\vk) & = 
  U^{\dagger} H_{d-1}(-\vk)^*  U,
  \nonumber \\ &=
  -(\sigma_3 U)^{\dagger} H_{d-1}(-\vk)^* (\sigma_3 U),\\
  H_{d-1}(\vk) &= U_\mathcal{R}^\dagger H_{d-1}(R\vk)U_\mathcal{R},\\
  H_{d-1}(\vk) &= U_\mathcal{CR}^\dagger H_{d-1}(R\vk)U_\mathcal{CR},
  \label{eq:Tred}
\end{align}
with
\begin{align}
  U &= \begin{pmatrix}
    0 & V_{{\cal T}}^{*} \\
    Q_{{\cal T}}^{*} & 0
  \end{pmatrix},\\
  U_\mathcal{R} &=
  \begin{pmatrix}
    Q_\mathcal{R} & 0\\
    0 & V_\mathcal{R}
  \end{pmatrix},\\
  U_\mathcal{CR} &=
  \begin{pmatrix}
    0 & Q_\mathcal{CR} \\
    V_\mathcal{CR} & 0
  \end{pmatrix}.
\end{align}
Here the first constraint has form of a time-reversal symmetry squaring to
$\mathcal{T}^2$ and a particle-hole symmetry squaring to $-\mathcal{T}^2$, while
the second (third) constraint has form of reflection symmetry with the same
(different) algebraic relation to $\mathcal{T}$ and $\mathcal{P}$ given by
$\sigma_\mathcal{T}$ ($\sigma_\mathcal{T}$ and $-\sigma_\mathcal{T}$,
respectively). 

({\em iii}).--- If $H_d$ has particle-hole symmetry but no time-reversal
symmetry, one has
\begin{align}
  H_{d-1}(\vk) & = 
  {\cal P}^2
  U^{\dagger} H_{d-1}(-\vk)^*  U,
  \nonumber \\ &=
  - {\cal P}^2
  (\sigma_3 U)^{\dagger} H_{d-1}(-\vk)^* (\sigma_3 U),\\
  H_{d-1}(\vk) &= U_\mathcal{R}^\dagger H_{d-1}(R\vk)U_\mathcal{R},\\
  H_{d-1}(\vk) &= U_\mathcal{CR}^\dagger H_{d-1}(R\vk)U_\mathcal{CR},
  \label{eq:Pred}
\end{align}
with
\begin{align}
  U_{\cal R} &= \begin{pmatrix}
    Q_{{\cal CR}} & 0 \\
    0 & V_{{\cal CR}} 
  \end{pmatrix},\\
  U_\mathcal{R} &=
  \begin{pmatrix}
    Q_\mathcal{R} & 0\\
    0 & V_\mathcal{R}
  \end{pmatrix},\\
  U_\mathcal{CR} &=
  \begin{pmatrix}
    0 & Q_\mathcal{CR} \\
    V_\mathcal{CR} & 0
  \end{pmatrix}.
\end{align}
In this case the first constraint has form of a time-reversal symmetry squaring
to $\mathcal{P}^2$ and a particle-hole symmetry squaring to $\mathcal{P}^2$,
while the second (third) constraint has form of reflection symmetry with the
same (different) algebraic relation to $\mathcal{T}$ and $\mathcal{P}$ given by
$\sigma_\mathcal{P}$ ($\sigma_\mathcal{P}$ and $-\sigma_\mathcal{P}$,
respectively). 

\section{Alternative derivation of Bott clock}
\label{app:K}

It is possible to establish the Bott clock structure of the classification tables \ref{tab:AZcomplex} and \ref{tab:AZreal} without the use of the reflection-matrix-based dimensional reduction scheme. Hereto we again consider the generalization of the exact sequence (\ref{eq:exactseq}) to $d>1$. We first note that the spaces $\cH{0}$ and $\cR{0}$ now contain $d-1$ dimensional gapped Hamiltonians $H(\vk')$, where the $d-1$ dimensional wavevector $\vk' = (k_1,k_2,\ldots,k_{d-1})$ contains the directions {\em parallel} to the reflection hyperplane. By combining reflection symmetry and time-reversal or particle-hole symmetry, the antiunitary symmetries that defined the spaces $\cH{0}$ and $\cR{0}$ now act as involutions linking $H(\vk')$ to $H(-\vk')^*$. The space of such functions, with the additional condition that $H(\vk')$ be ``trivial'' on the (hyper)planes $k_{\alpha} = 0$, $\alpha=1,2,\ldots,d-1$, is denoted $\bar \Omega^{d-1} \mathcal{C}_n$ and $\bar \Omega^{d-1} \mathcal{R}_n$, for the complex and real classes, respectively, where, instead of the Cartan labels, we use the symbols $\mathcal{C}_n$ and $\mathcal{R}_n$ to refer to the Altland-Zirnbauer classes, see Table \ref{tab:classes}. Hence, the proper generalization of the Hamiltonian spaces $\cH{0}$ and $\cR{0}$ to the $d$-dimensional case are the spaces $\bar \Omega^{d-1} \cH{0}$ and $\bar \Omega^{d-1} \cR{0}$. The corresponding generalization of the exact sequence (\ref{eq:exactseq}) then reads
\begin{align}
  \pi_1(\bar \Omega^{d-1} \cR{0}) &\overset{i_1}{\hookrightarrow}\pi_1(\bar \Omega^{d-1}\cH{0})\overset{j_1}{\hookrightarrow}\pi_1(\bar \Omega^{d-1}\cH{0},\bar \Omega^{d-1}\cR{0})
  \nonumber \\ &\overset{\delta}{\rightarrow}\pi_0(\bar \Omega^{d-1}\cR{0})\overset{i_0}{\hookrightarrow}\pi_0(\bar \Omega^{d-1}\cH{0}).
  \label{eq:exactseqd}
\end{align}  

A central result in $K$ theory is that the zeroth and first homotopy groups appearing in the exact sequence (\ref{eq:exactseq}) satisfy a periodicity rule,
\begin{align}
  \pi_m(\bar \Omega^{d} \mathcal{C}_n) &= \pi_m(\mathcal{C}_{n-d \mod\, 2}),\nonumber \\
  \pi_m(\bar \Omega^{d} \mathcal{R}_n) &= \pi_m(\mathcal{R}_{n-d \mod\, 8}).
  \label{eq:theorem}
\end{align}
This allows one to directly map the exact sequence for the $d$-dimensional case to the exact sequence for the one-dimensional case. Since the assignment of the spaces $\cH{0}$ and $\cR{0}$ follows the Bott clock structure, see Tables \ref{tab:complex} and \ref{tab:real}, the periodicty implied by Eq.\ (\ref{eq:theorem}) immediately extends the Bott clock structure to arbitrary dimensions larger than one.

\section{Topological classification for $d=1$}\label{app:exactsec}

\label{app:exactseq}

In Tables~\ref{tab:complex} and \ref{tab:real} we give the Cartan labels for the spaces $\cH{0}$ and $\cR{0}$, as well as the corresponding exact sequences for each of the 36 symmetry classes with reflection and/or $\mathcal{CR}$ symmetry. Below we give details for a few special cases that need additional considerations to resolve the sequence and that were not considered in the main text.

\begin{table}
  \begin{tabular}{|c|c|c|c|}
    \hline
    Class & $\cH{0}$ & $\cR{0}$ & Exact sequence\\\hline
    A$^{\mathcal{R}}$ & A & A$^2$ & $0\overset{i_1}{\hookrightarrow}0\overset{j_1}{\hookrightarrow}\bZ\overset{\delta}{\rightarrow}\bZ^2\overset{i_0}{\hookrightarrow}\bZ$\\\hline
    AIII$^{\mathcal{R}_+}$ & AIII & AIII$^2$ & $\bZ^2\overset{i_1}{\hookrightarrow}\bZ\overset{j_1}{\hookrightarrow}0\overset{\delta}{\rightarrow}0\overset{i_0}{\hookrightarrow}0$\\\hline
    \hline
    A$^{\mathcal{CR}}$ & A & AIII & $\bZ\overset{i_1}{\hookrightarrow}0\overset{j_1}{\hookrightarrow}0\overset{\delta}{\rightarrow}0\overset{i_0}{\hookrightarrow}0$\\\hline
    AIII$^{\mathcal{R}_-}$ & AIII & A  & $0\overset{i_1}{\hookrightarrow}\bZ\overset{j_1}{\hookrightarrow}\bZ^2\overset{\delta}{\rightarrow}\bZ\overset{i_0}{\hookrightarrow}0$\\\hline
  \end{tabular}
\caption{The complete list of the exact sequences for $d=1$ complex class
topological insulators with reflection symmetry. The table consists of two
subtables, within each subtable, the classifying spaces $\cH{0}$ and
$\cR{0}$ run along the Bott clock.}
\label{tab:complex}
\end{table}

\begin{table}
  \begin{tabular}{|c|c|c|c|}
    \hline
    Class & $\cH{0}$ & $\cR{0}$ & Exact sequence\\\hline
    AI$^{\mathcal{R}_+}$ & AI & AI$^2$& $\bZ_2^2\overset{i_1}{\hookrightarrow}\bZ_2\overset{j_1}{\hookrightarrow}\bZ\overset{\delta}{\rightarrow}\bZ^2\overset{i_0}{\hookrightarrow}\bZ$\\\hline
    BDI$^{\mathcal{R}_{++}}$ & BDI & BDI$^2$ & $\bZ_2^2\overset{i_1}{\hookrightarrow}\bZ_2\overset{j_1}{\hookrightarrow}\bZ_2\overset{\delta}{\rightarrow}\bZ^2_2\overset{i_0}{\hookrightarrow}\bZ_2$\\\hline
    D$^{\mathcal{R}_+}$ & D & D$^2$ & $0\overset{i_1}{\hookrightarrow}0\overset{j_1}{\hookrightarrow}\bZ_2\overset{\delta}{\rightarrow}\bZ^2_2\overset{i_0}{\hookrightarrow}\bZ_2$\\\hline
    DIII$^{\mathcal{R}_{++}}$ & DIII & DIII$^2$ & $2\bZ^2\overset{i_1}{\hookrightarrow}2\bZ\overset{j_1}{\hookrightarrow}0\overset{\delta}{\rightarrow}0\overset{i_0}{\hookrightarrow}0$\\\hline
    AII$^{\mathcal{R}_{+}}$ & AII & AII$^2$ & $0\overset{i_1}{\hookrightarrow}0\overset{j_1}{\hookrightarrow}2\bZ\overset{\delta}{\rightarrow}2\bZ^2\overset{i_0}{\hookrightarrow}2\bZ$\\\hline
    CII$^{\mathcal{R}_{++}}$ & CII & CII$^2$ & $0\overset{i_1}{\hookrightarrow}0\overset{j_1}{\hookrightarrow}0\overset{\delta}{\rightarrow}\bZ^2\overset{i_0}{\hookrightarrow}\bZ$\\\hline
    C$^{\mathcal{R}_+}$ & C & C$^2$& $0\overset{i_1}{\hookrightarrow}0\overset{j_1}{\hookrightarrow}0\overset{\delta}{\rightarrow}0\overset{i_0}{\hookrightarrow}0$\\\hline
    CI$^{\mathcal{R}_{++}}$ & CI & CI$^2$ & $\bZ^2\overset{i_1}{\hookrightarrow}\bZ\overset{j_1}{\hookrightarrow}0\overset{\delta}{\rightarrow}0\overset{i_0}{\hookrightarrow}0$\\\hline
    \hline
    AI$^{\mathcal{R}_-}$ & AII & A & $0\overset{i_1}{\hookrightarrow}0\overset{j_1}{\hookrightarrow}0\overset{\delta}{\rightarrow}\bZ\overset{i_0}{\hookrightarrow}2\bZ$\\\hline
    BDI$^{\mathcal{R}_{--}}$ & CII & AIII & $\bZ\overset{i_1}{\hookrightarrow}0\overset{j_1}{\hookrightarrow}0\overset{\delta}{\rightarrow}0\overset{i_0}{\hookrightarrow}0$\\\hline
    D$^{\mathcal{R}_-}$ & C & A & $0\overset{i_1}{\hookrightarrow}0\overset{j_1}{\hookrightarrow}\bZ\overset{\delta}{\rightarrow}\bZ\overset{i_0}{\hookrightarrow}0$\\\hline
    DIII$^{\mathcal{R}_{--}}$ & CI & AIII & $\bZ\overset{i_1}{\hookrightarrow}\bZ\overset{j_1}{\hookrightarrow}\bZ_2\overset{\delta}{\rightarrow}0\overset{i_0}{\hookrightarrow}0$\\\hline
    AII$^{\mathcal{R}_{-}}$ & AI & A & $0\overset{i_1}{\hookrightarrow}\bZ_2\overset{j_1}{\hookrightarrow}\bZ_2\overset{\delta}{\rightarrow}\bZ\overset{i_0}{\hookrightarrow}\bZ$\\\hline
    CII$^{\mathcal{R}_{--}}$ & BDI & AIII & $\bZ\overset{i_1}{\hookrightarrow}\bZ_2\overset{j_1}{\hookrightarrow}0\overset{\delta}{\rightarrow}0\overset{i_0}{\hookrightarrow}\bZ_2$\\\hline
    C$^{\mathcal{R}_-}$ & D & A & $0\overset{i_1}{\hookrightarrow}0\overset{j_1}{\hookrightarrow}2\bZ\overset{\delta}{\rightarrow}\bZ\overset{i_0}{\hookrightarrow}\bZ_2$\\\hline
    CI$^{\mathcal{R}_{--}}$ & DIII & AIII & $\bZ\overset{i_1}{\hookrightarrow}2\bZ\overset{j_1}{\hookrightarrow}0\overset{\delta}{\rightarrow}0\overset{i_0}{\hookrightarrow}0$\\\hline
    \hline
    AI$^{\mathcal{CR}_-}$ & C & CI & $\bZ\overset{i_1}{\hookrightarrow}0\overset{j_1}{\hookrightarrow}0\overset{\delta}{\rightarrow}0\overset{i_0}{\hookrightarrow}0$\\\hline
    BDI$^{\mathcal{R}_{+-}}$ & CI & AI & $\bZ_2\overset{i_1}{\hookrightarrow}\bZ\overset{j_1}{\hookrightarrow}\bZ^2\overset{\delta}{\rightarrow}\bZ\overset{i_0}{\hookrightarrow}0$\\\hline
    D$^{\mathcal{CR}_{+}}$ & AI & BDI & $\bZ_2\overset{i_1}{\hookrightarrow}\bZ_2\overset{j_1}{\hookrightarrow}\bZ_2^2\overset{\delta}{\rightarrow}\bZ_2\overset{i_0}{\hookrightarrow}\bZ$\\\hline
    DIII$^{\mathcal{R}_{-+}}$ & BDI & D & $0\overset{i_1}{\hookrightarrow}\bZ_2\overset{j_1}{\hookrightarrow}\bZ_2^2\overset{\delta}{\rightarrow}\bZ_2\overset{i_0}{\hookrightarrow}\bZ_2$\\\hline
    AII$^{\mathcal{CR}_-}$ & D & DIII & $2\bZ\overset{i_1}{\hookrightarrow}0\overset{j_1}{\hookrightarrow}0\overset{\delta}{\rightarrow}0\overset{i_0}{\hookrightarrow}\bZ_2$\\\hline
    CII$^{\mathcal{R}_{+-}}$ & DIII & AII & $0\overset{i_1}{\hookrightarrow}2\bZ\overset{j_1}{\hookrightarrow}2\bZ^2\overset{\delta}{\rightarrow}2\bZ\overset{i_0}{\hookrightarrow}0$\\\hline
    C$^{\mathcal{CR}_+}$ & AII & CII & $0\overset{i_1}{\hookrightarrow}0\overset{j_1}{\hookrightarrow}0\overset{\delta}{\rightarrow}0\overset{i_0}{\hookrightarrow}0$\\\hline
    CI$^{\mathcal{R}_{-+}}$ & CII & C & $0\overset{i_1}{\hookrightarrow}0\overset{j_1}{\hookrightarrow}0\overset{\delta}{\rightarrow}0\overset{i_0}{\hookrightarrow}0$\\\hline
    \hline
    AI$^{\mathcal{CR}_+}$ & D & BDI & $\bZ_2\overset{i_1}{\hookrightarrow}0\overset{j_1}{\hookrightarrow}0\overset{\delta}{\rightarrow}\bZ_2\overset{i_0}{\hookrightarrow}\bZ_2$\\\hline
    BDI$^{\mathcal{R}_{-+}}$ & DIII & D & $0\overset{i_1}{\hookrightarrow}2\bZ\overset{j_1}{\hookrightarrow}\bZ\overset{\delta}{\rightarrow}\bZ_2\overset{i_0}{\hookrightarrow}0$\\\hline
    D$^{\mathcal{CR}_-}$ & AII & DIII & $2\bZ\overset{i_1}{\hookrightarrow}0\overset{j_1}{\hookrightarrow}0\overset{\delta}{\rightarrow}0\overset{i_0}{\hookrightarrow}2\bZ$\\\hline
    DIII$^{\mathcal{R}_{+-}}$ & CII & AII & $0\overset{i_1}{\hookrightarrow}0\overset{j_1}{\hookrightarrow}2\bZ\overset{\delta}{\rightarrow}2\bZ\overset{i_0}{\hookrightarrow}0$\\\hline
    AII$^{\mathcal{CR}_+}$ & C & CII & $0\overset{i_1}{\hookrightarrow}0\overset{j_1}{\hookrightarrow}0\overset{\delta}{\rightarrow}0\overset{i_0}{\hookrightarrow}0$\\\hline
    CII$^{\mathcal{R}_{-+}}$ & CI & C & $0\overset{i_1}{\hookrightarrow}\bZ\overset{j_1}{\hookrightarrow}\bZ\overset{\delta}{\rightarrow}0\overset{i_0}{\hookrightarrow}0$\\\hline
    C$^{\mathcal{CR}_-}$ & AI & CI & $\bZ\overset{i_1}{\hookrightarrow}\bZ_2\overset{j_1}{\hookrightarrow}0\overset{\delta}{\rightarrow}0\overset{i_0}{\hookrightarrow}\bZ$\\\hline
    CI$^{\mathcal{R}_{+-}}$ & BDI & AI & $\bZ_2\overset{i_1}{\hookrightarrow}\bZ_2\overset{j_1}{\hookrightarrow}2\bZ\overset{\delta}{\rightarrow}\bZ\overset{i_0}{\hookrightarrow}\bZ_2$\\\hline
  \end{tabular}
\caption{The complete list of the exact sequences for $d=1$ real class
topological insulators and superconductors with reflection symmetry. The table
consists of four subtables, within each subtable, the classifying spaces $\cH{0}$
and $\cR{0}$ run along the Bott clock.}
\label{tab:real}
\end{table}

\subsubsection{The class AIII$^{\mathcal{R}_{-}}$ in $d=1$}\label{sec:AIIIRm}

Following Subsection \ref{sec:exactseq} we write $H(k) = H(0) \oplus H'(k)$, where $H'(0)$ is topologically trivial, and use the method of Subsection \ref{sec:exactseq} to classify the Hamiltonians $H'(k)$. Without loss of generality we may choose the unitary operators encoding chiral and reflection operations as $U_{\cal C} = \sigma_3$, $U_{\cal R} = \sigma_1$, which ensures that ${\cal C}$ and ${\cal R}$ anticommute. With this choice, chiral symmetry implies that $H'(k)$ has the form\cite{1367-2630-12-6-065010}
\begin{equation}
  H'(k) = \begin{pmatrix}
  0 & r'(k) \\ r'(k)^{\dagger} & 0
  \end{pmatrix},
\label{eq:Hkr}
\end{equation}
whereas reflection symmetry imposes the further constraint $r'(k) = r'(-k)^{\dagger}$. In particular, the blocks $r'(0)$ and $r'(\pi)$ at the reflection-symmetric momemta $k=0$, $\pi$ are hermitian. The numbers $N'(0)$ and $N'(\pi)$ of negative eigenvalues of $r'(0)$ or $r'(\pi)$ determines the number of occupied states at $k=0$, $\pi$ that are {\em even} under reflection. We (arbitrarily) take the condition that $H'(0)$ be trivial to imply that $N'(0) = 0$.

In the language of Sec.\ \ref{sec:exactseq}, we thus see that $\cH{0}$ is the space of Hamiltonians in symmetry class AIII, whereas $\cR{0}$ is class A. Taking the zeroth and first fundamental groups of these two classes from Table \ref{tab:classes}, we find that the corresponding exact sequence (\ref{eq:exactseq}) reads
\begin{align}
  0\overset{i_1}{\hookrightarrow}\bZ\overset{j_1}{\hookrightarrow}\bZ\rtimes\bZ\overset{\delta}{\rightarrow}\bZ\overset{i_0}{\hookrightarrow}0.
  \label{eq:a3mexact}
\end{align}
We conclude that $\pi_1(\cH{0},\cR{0})=\bZ\rtimes\bZ = \bZ^2$. 

The left and the right part of the exact sequence (\ref{eq:a3mexact}) each contribute each a single topological index. We identify the first index $N'$, which comes from the right part of the exact sequence, as the number $N'(\pi)$ of even occupied states (under reflection symmetry) at the reflection symmetric point $k=\pi$. (Recall that by assumption $N'(0) = 0$.) The second topological index $W'$ enumerates the equivalence classes of loops in AIII, and can be calculated as twice the winding number of $\det r'(k)$ between $k = 0$ and $k=\pi$. [Note that $\det r'(0) = 1$, since $H'(0)$ is trivial, whereas $\det r'(\pi) = \pm 1$ and $\det r'(k) = \det r'(-k)^*$.] Since an odd value of $W'$ implies that odd parity of $N'$ and vice versa, the allowed values of $(N',W')$ are subject to the constraint that $N' + W'$ be even. 

Returning to the original Hamiltonian $H(k) = H(0) \oplus H'(k)$, we see that the addition of the $k$-independent part $H(0)$ does not affect the index $W'$, whereas the first topological index has to be calculated as $N' = N(\pi) - N(0)$, {\em i.e.}, as the difference of the number of occupied even states at $k=\pi$ and $k=0$. To see that all allowed values for the two topological indices $N'$ and $W'$ are indeed attained, we construct generating Hamiltonians with $(N',W) = (1,\pm 1)$,
\begin{eqnarray}
  H^{(1,\pm 1)}(k) = \cos k \sigma_1 \mp \sin k \sigma_2.
  \label{eq:Hpma3}
\end{eqnarray}


\subsubsection{The class BDI$^{\mathcal{R}_{+-}}$}

At each $k$-point the Hamiltonian satisfies $\mathcal{RT}$ and $\mathcal{RP}$ symmetries with $(\mathcal{RT})^2=1$, $(\mathcal{RP})^2=-1$, thus $\cH{0}$ is the space of Hamiltonians in symmetry class CI. Without loss of generality we take a representation in which $\mathcal{RT} = K$, $\mathcal{RP} = \sigma_2 K$, so that generic $H(k)$ can be written in the form
\begin{equation}
  H(k) = \begin{pmatrix} h(k) & d(k) \\ d(k) & -h(k) \end{pmatrix},
  \label{eq:HCI}
\end{equation}
with real and symmetric $h(k)$ and $d(k)$. Since $\mathcal{R}$ anticommutes with $\mathcal{P}$ and, hence, with $\mathcal{RP}$, but commutes with $\mathcal{T}$ and $\mathcal{RT}$ we may then use the unitary matrix $U_{\cal R} = \sigma_3$ to represent the reflection operation. This implies $d(k) = -d(-k)$ and, in particular, $d(0) = d(\pi) = 0$, so that $\cR{0}$ is the space of class AI Hamiltonians. As in the previous Subsection we write $H(k) = H(0) \oplus H'(k)$, where $H'(0)$ is trivial, where we (again, arbitrarily) define ``trivial'' as all occupied states being odd under reflection. The exact sequence (\ref{eq:exactseq}) then takes the form
\begin{align}
  \bZ_2\overset{i_1}{\hookrightarrow}\bZ\overset{j_1}{\hookrightarrow}\bZ\rtimes\bZ\overset{\delta}{\rightarrow}\bZ\overset{i_0}{\hookrightarrow}0,
  \label{eq:bdipmexact}
\end{align}
leading to $\pi_1(\cH{0},\cR{0})=\bZ^2$. We identify the two topological invariants $N'$ and $W'$ as the number of negative eigenvalues of $h'(\pi)$ and as twice the winding number of $h(k) + i d(k)$ between $k=0$ and $k=\pi$, respectively. For the original Hamiltonian $H(k)$, the first invariant $N'$ becomes the difference of the number of negative eigenvalues of $h(k)$ at $k=\pi$ and $k=0$. As in the previous case, the topological numbers are subject to the constraint that $N' + W'$ be even. The two generators are
\begin{align}
  H^{(1,\pm 1)}(k)&=\cos k\sigma_3 \pm \sin k\sigma_1.
  \label{eq:Hbdipm}
\end{align}

\subsubsection{The class CII$^{\mathcal{R}_{+-}}$}

Since $\mathcal{R}$ anticommutes with $\mathcal{P}$ and $\mathcal{P}^2 = -1$ one has $(\mathcal{R P})^2 = 1$. Similarly, since $\mathcal{R}$ commutes with $\mathcal{T}$ and $\mathcal{T}^2 = -1$ one has $(\mathcal{R T})^2 = -1$. This motivates the representation $\mathcal{RT}=\tau_2 \sigma_3K$, $\mathcal{RP}=\sigma_2 K$, $\mathcal{R}=\sigma_3$. To get an explicit representation of matrices $H(k)$ for $0 < k < \pi$ it is advantageous to define two alternative sets of Pauli matrices, $\mu_1 = \sigma_1$, $\mu_2 = \tau_2 \sigma_3$, $\mu_3 = -\tau_2 \sigma_2$, $\lambda_1 = \tau_1 \sigma_1$, $\lambda_2 = \tau_2$, $\lambda_3 = \tau_3 \sigma_1$, so that $\mathcal{RP} = \mu_1 K$, $\mathcal{RT} = \mu_2 K$. Using the $2 \times 2$ matrix structure corresponding to the $\mu$ matrices, we find that $H(k)$ for $0 < k < \pi$ has the form
\begin{equation}
  H(k) = \begin{pmatrix} 0 & r(k) \\ r(k)^{\dagger} & 0 \end{pmatrix},
  \label{eq:HKRRR}
\end{equation}
with $r(k)$ complex antisymmetric and $r(k) = \lambda_2 r^*(-k) \lambda_2$. We conclude that $\cH{0}$ is the space of gapped Hamiltonians in class DIII. To identify the symmetry of $H(k)$ at the symmetric momenta $k=0$, $\pi$ we use the original representation in terms of Pauli matrices $\sigma$, $\tau$, and find that $H(k)$ is of the form
\begin{equation}
  H(k) = \begin{pmatrix} h & 0 \\ 0 & h^* \end{pmatrix},\ \
  k=0,\pi,
\end{equation}
with $h = \tau_2 h^* \tau_2$, which manifestly places $\cR{0}$ is in class AII. This gives the exact sequence
\begin{align}
  0\overset{i_1}{\hookrightarrow}2\bZ\overset{j_1}{\hookrightarrow}2\bZ\times2\bZ\overset{\delta}{\rightarrow}2\bZ\overset{i_0}{\hookrightarrow}0.
  \label{eq:c2pmexact}
\end{align}
The topological indices $N'$ and $W'$ are the difference in the number of occupied odd states at $k=0$ and $k=\pi$ ({\em i.e.}, the difference of the number of negative eigenvalues of $h(k)$ for $k=0$ and $\pi$) and the winding number of $\mbox{det}\, r(k)$ for $k$ between $0$ and $\pi$, respectively. (Note that the condition $r(k) = \lambda_2 r^*(-k) \lambda_2$ for $k=0$, $\pi$ implies that the Pfaffians of $r(0)$ and $r(\pi)$ are real, which implies that $\mbox{det}\, r(k)$ has an integer winding number for $k$ between $0$ and $\pi$.) They are subject to the constraint that $N' + W'$ be even. Generators are
\begin{align}
  H^{(1,\pm 1)}(k)&=\tau_2(\sigma_1\cos k\pm\sin k\sigma_3).
  \label{eq:Hc2mp}
\end{align}

\subsubsection{The class D$^{\mathcal{CR}_+}$}

The combined symmetry $\mathcal{P}\,\mathcal{CR}$ serves as an effective time-reversal symmetry at each $k$ point squaring to one, so that $\cH{0}$ is the class AI. At the reflection symmetric points the $\mathcal{CR}$ symmetry contributes as an additionally chiral symmetry that commutes with $\mathcal{PCR}$, so that $\cR{0}$ is the class BDI. This gives the exact sequence
\begin{align}
  \bZ_2\overset{i_1}{\hookrightarrow}\bZ_2\overset{j_1}{\hookrightarrow}\bZ_2\times\bZ_2\overset{\delta}{\rightarrow}\bZ_2\overset{i_0}{\hookrightarrow}\bZ.
  \label{eq:dcrpexact}
\end{align}
To see that $\pi_1(\cH{0},\cR{0}) = \bZ_2 \times \bZ_2$ we first need to verify that the image of $i_1$ is the identity element in $\pi_1(\cH{0})$. Hereto, we choose the representation $\mathcal{PCR} = K$ and $\mathcal{CR} = \tau_3$ and show that that the generator of $\pi_1(\cR{0})$, 
\begin{align}
  H(k)&=\tau_1(\sigma_3\cos k+\sigma_1\sin k),
  \label{eq:bdiloop}
\end{align}
is mapped to the trivial element in $\pi_1(\cH{0})$. This is indeed the case, since $H(k)$ is the diagonal sum of two nontrivial Hamiltonians in AI.

To find the group structure of $\pi_1(\cH{0},\cR{0})$ we need to inspect the topological indices $N'$ and $W'$ coming from the right and left parts of the exact sequence. We use the fact that at $k=0,\pi$ the Hamiltonian $H$ has the standard form
\begin{equation}
  H(k) = \begin{pmatrix} 0 & o(k) \\ o(k)^{\rm T} & 0 \end{pmatrix},\ \
  k=0,\, \pi,
\end{equation}
where we take $o(k)$ to be orthogonal, $k=0$, $\pi$. (Orthogonal blocks $o(k)$ for $k=0$, $\pi$ can always be achieved by continuous deformation of $H(k)$.) This gives a reference basis for the occupied states given by the vectors
\begin{equation}
  \vert u_{{\rm ref},j}(k) \rangle = \begin{pmatrix} o_j(k) \\
  - e_j \end{pmatrix},\ \ j=1,2,\ldots,N,
\end{equation}
where $o_j(k)$ is the $j$th column of the orthogonal matrix $o(k)$, $e_j$ the
$j$th unit vector, and $2N$ the dimension of $H(k)$. We use $\vert u_{j}(k)
\rangle$ to denote a basis for the occupied states, continuous as a function of
$k$ and such that $\vert u_{j}(0) \rangle = \vert u_{{\rm ref},j}(0) \rangle$.
We can then define the topological indices $N' = \mbox{sign}\, \det o(0) \det
o(\pi)$ and $W' = \mbox{sign}\, \det [\langle u_{i}(\pi) \vert u_{{\rm
ref},j}(\pi)\rangle]$. The corresponding generators are
\begin{align}
  H^{(-1,1)}(k) &= \tau_1 \cos k + \tau_3 \sin k, \nonumber \\
  H^{(1,-1)}(k) &= \tau_1 \cos(2 k) + \tau_3 \sin( 2 k).
\end{align}
To see, note that $H^{(-1,1)}(k)$ has has $\vert u_{\rm ref}(0) \rangle = (1,-1)/\sqrt{2}$, $\vert u_{\rm ref}(\pi) \rangle = (-1,-1)/\sqrt{2}$, and $\vert u(k) \rangle = (\cos(k/2)-\sin(k/2),-\cos(k/2)-\sin(k/2))/\sqrt{2}$, so that $W' = 1$. Similarly, $H^{(1,-1)}(k)$ has $\vert u_{\rm ref}(0) \rangle = \vert u_{\rm ref}(\pi) \rangle = (1,-1)/\sqrt{2}$, and $\vert u(k) \rangle = (\cos(k)-\sin(k),-\cos(k)-\sin(k))/\sqrt{2}$, which gives $W' = -1$. The $\bZ_2 \times \bZ_2$ group structure follows upon verifying that the diagonal sum of each generator with itself has trivial indices. 

\subsubsection{The class $DIII^{\mathcal{R}_{-+}}$}

In this case $\mathcal{TR}$ and $\mathcal{PR}$ give effective time-reversal and particle-hole symmetries that both square to $1$. We may take the representation $\mathcal{TR} = K$, $\mathcal{PR} = \sigma_3 K$, and $\mathcal{R} = \sigma_2$ and find that $H(k)$ is of the form
\begin{equation}
  H(k) = \begin{pmatrix} 0 & r(k) \\ r(k)^{\rm T} & 0 \end{pmatrix},
  \label{eq:HDIIIR}
\end{equation}
with $r(k)$ real and $r(k) = -r(-k)^{\rm T}$. Hence we have $\cH{0}$ as class BDI and $\cR{0}$ as class D. The exact sequence (\ref{eq:exactseq}) takes the form
\begin{align}
  0 \overset{i_1}{\hookrightarrow}\bZ_2\overset{j_1}{\hookrightarrow}\bZ_2\times\bZ_2\overset{\delta}{\rightarrow}\bZ_2\overset{i_0}{\hookrightarrow}\bZ_2,
\end{align}
so that $\pi_1(\cH{0},\cR{0}) = \bZ_2 \times \bZ_2$. To verify, we note that the image of $i_0$ is trivial, since $\det r(k)$ is always positive at $k=0$, $\pi$. The topological invariants are $N' = \mbox{sign}\, [\mbox{Pf} r(0)\, \mbox{Pf} r(\pi)]$ and ``winding number'' $W'$ of the real matrices $r(k)$ upon taking $k$ from $0$ to $\pi$, which is the parity of the number twofold degenerate crossings of the (complex) eigenvalues of $r(k)$ on the negative real axis. The generators are of the from (\ref{eq:HDIIIR}) with
\begin{align}
  r^{(-1,1)}(k) &= i \tau_2 \cos k + \sin k, \nonumber \\
  r^{(1,-1)}(k) &= i \tau_2 \cos (2 k) + \sin (2 k).
\end{align}
One verifies that the diagonal sum of each generator with itself gives a trivial element, so that the group structure is indeed $\bZ_2 \times \bZ_2$.

\subsubsection{The class BDI$^{\mathcal{R}_{-+}}$}

For this combination of symmetries we have $(\mathcal{RT})^2=-1$, $(\mathcal{RP})^2=1$, so that $\cH{0}$ belongs to the class DIII. We choose the explicit representations $\mathcal{RT} = \sigma_2 K$, $\mathcal{RP} = K$, and $\mathcal{R} = \sigma_3$. The Hamiltonian $H(k)$ can then be cast into the form 
\begin{equation}
  H(k) = i \begin{pmatrix} 
  h(k) & d(k) \\ d(k) & -h(k)
  \end{pmatrix},
  \label{eq:HDIII}
\end{equation}
where $h(k)$ and $d(k)$ are real and antisymmetric and $d(k) = -d(-k)$. We see that $\cH{0}$ is in class DIII, whereas $\cR{0}$ is in class D. We again write $H(k) = H(0) \oplus H'(k)$ with $H'(0)$ trivial, calling $H'(0)$ ``trivial'' if the Pfaffian $\mbox{Pf}\, h'(0)$ is positive. The exact sequence (\ref{eq:exactseq}) then reads
\begin{align}
  0\overset{i_1}{\hookrightarrow}2\bZ\overset{j_1}{\hookrightarrow} \bZ \overset{\delta}{\rightarrow}\bZ_2\overset{i_0}{\hookrightarrow}0.
  \label{eq:bdimpexact}
\end{align}

To see that the exact sequence (\ref{eq:bdimpexact}) indeed gives $\pi_1(\cH{0},\cR{0})= \bZ$, we identify the topological indices $N'$ and $W'$ coming from the right and left parts of the exact sequence. In the present case the index $N' = \mbox{Pf}\, h(\pi)$, which is a $\bZ_2$ index, not an integer (as in the previous two examples). The index $W'$ is twice the winding number of the {\em Pfaffian} $\mbox{Pf}\, [h(k) + i d(k)]$ for $k$ between $0$ and $\pi$. As in the previous two examples we have the constraint that $N' + W'$ be even. Since $N'$ is a $\bZ_2$ index, this implies that the integer $W'$ alone is sufficient to determine the topological classification of $H'(k)$. Since $H(k)$ and $H'(k)$ have the same winding numbers, the same topological index applies to the full Hamiltonian $H(k)$. As a generator we can take the four-band Hamiltonian
\begin{equation}
  H(k) = \begin{pmatrix} \tau_2 \cos k & \tau_2 \sin k \\ \tau_2 \sin k & - \tau_2 \cos k \end{pmatrix}.
\end{equation}

\subsubsection{The class DIII$^{\mathcal{R}_{--}}$}

Here we choose a representation $\mathcal{T}=\tau_2K$, $\mathcal{P}=\tau_2\sigma_2K$ and $\mathcal{R}=\tau_2$, for which $H(k)$ is of the form (\ref{eq:HCI}) with real and symmetric matrices $h(k)$ and $d(k)$ satisfying $h(k) = \tau_2 h(-k) \tau_2$, $d(k) = \tau_2 d(-k) \tau_2$. The commutation with $\tau_2$ induces a two-dimensional representation of the complex numbers for $k=0$, $\pi$, see the discussion following Eq.\ (\ref{eq:complex}), so that we conclude that $\cH{0}$ and $\cR{0}$ are classes CI and AIII, respectively. The exact sequence (\ref{eq:exactseq}) then reads
\begin{align}
  \bZ\overset{i_1}{\hookrightarrow}\bZ\overset{j_1}{\hookrightarrow}\bZ_2\overset{\delta}{\rightarrow}0\overset{i_0}{\hookrightarrow}0.
\end{align}
To show that $\pi_1(\cH{0},\cR{0}) = \bZ_2$, we need to show that the image of $i_1$ is $2\bZ$. To this end it is sufficient to consider the Hamiltonian
\begin{align}
  H(k)&=\sigma_3\cos k+\sigma_1\sin k,
  \label{eq:d3rmm}
\end{align}
which is a generator for the loop space of AIII. Upon inclusion the loop space of CI this Hamiltonian is mapped to $i_1[H(k)] = H(k) \otimes \tau_0$, which is not the generator of the CI loop space. Instead, the generator of the CI loop space is $H(k) \oplus \sigma_3$, so that we conclude that the topological index of $i_1[H(k)]$ is equal to ``2''. The generator of the CI loop space is also the generator of $\pi_1(\cH{0},\cR{0})$.



\subsubsection{The class C$^{\mathcal{CR}_-}$}

For this symmetry class $\cH{0}$ is class AI, whereas $\cR{0}$ is class CI. We need to prove that the image of $i_1$ is whole of $\bZ_2$. To this end it is
enough to verify that the particular example
\begin{align}
  H(k)&=\sigma_3\cos k+\sigma_1\sin k,
  \label{eq:c1loop}
\end{align}
with symmetries $\mathcal{T}=K$ and $\mathcal{P}=\sigma_2K$, is an element of CI and is mapped to a nontrivial loop in AI. 

\subsubsection{The class CII$^{\mathcal{R}_{--}}$}

Here one has $\cH{0}$ is BDI and $\cR{0}$ is AIII.
Similarly to the previous case, we need to show that the image of $i_1$ is $\bZ_2$. Again, it is sufficient to find an example,
\begin{align}
  H(k)&=\sigma_1\cos k+\sigma_2\sin k,\ \ 0 \le k \le 2 \pi,
  \label{eq:c2mm}
\end{align}
with chiral symmetry given by $\mathcal{C}=\sigma_3$. One verifies that this example constitutes a loop in AIII. In order to identify the above Hamiltonian as a loop in BDI, one has to use the two-dimensional representation of the complex numbers, see the discussion following Eq.\ (\ref{eq:complex}), which turns $H(k)$ into a matrix of the form (\ref{eq:Hkr}) with 
\begin{equation}
  r(k) = \cos k - i \tau_2 \sin k,
\end{equation}
which is a nontrivial loop in BDI.

\subsubsection{The class CI$^{\mathcal{R}_{--}}$}

For this symmetry class we have $(\mathcal{RP})^2 = 1$, $(\mathcal{RT})^2 = -1$, which motivates the representation $\mathcal{RP} = \sigma_1 K$, $\mathcal{RT} = \tau_3 \sigma_2 K$, and $\mathcal{R} = \sigma_3$. This implies that $\cH{0}$ is class DIII. To get an explicit representation of matrices $H(k)$ for $0 < k < \pi$ it is advantageous to define two alternative sets of Pauli matrices $\mu_1 = \sigma_1$, $\mu_2 = \tau_3 \sigma_2$, $\mu_3 = \tau_3 \sigma_3$, $\lambda_1 = \tau_1 \sigma_1$, $\lambda_2 = \tau_2 \sigma_1$, $\lambda_3 = \tau_3$, so that $\mathcal{RP} = \mu_1 K$, $\mathcal{RT} = \mu_2 K$. Using the $2 \times 2$ matrix structure corresponding to the $\mu$ matrices, we find that $H(k)$ for $0 < k < \pi$ has the form (\ref{eq:HKRRR})
with $r(k)$ complex antisymmetric. For $k=0$, $\pi$, it is more convenient to use the original Pauli matrices $\sigma$, $\tau$ and one finds that $H$ has the form
\begin{equation}
  H(k) = \begin{pmatrix} h & 0 \\ 0 & \tau_3 h^* \tau_3 \end{pmatrix},\ \
  k=0,\, \pi,
\end{equation}
where the hermitian $h$ satisfies the additional constraint $h = -\tau_3 h \tau_3$. This places $H(0)$ and $H(\pi)$ in class AIII.

In order to resolve the exact sequence, we have to show that the image of $i_1$ is all of $2\bZ$. Hereto we consider the example
\begin{align}
  h(k)&= \tau_1 \cos k+\tau_2 \sin k,\ \ 0 \le k \le 2 \pi,
  \label{eq:c1rmm}
\end{align}
which is a nontrivial ``loop'' in AIII. Seen as a loop in DIII, one has $r(k) = \lambda_2 \sin k + i \lambda_2 \cos k$, which indeed is the generator of the first homotopy group.

\subsubsection{The class CI$^{\mathcal{R}_{+-}}$}

We have $(\mathcal{RP})^2 = (\mathcal{RT})^2 = 1$, which motivates the
representation $\mathcal{RT} = K$, $\mathcal{RP} = \sigma_3 K$, $\mathcal{R} =
\sigma_1$, from which it follows that $H(k)$ is of the form (\ref{eq:HKRRR})
with $r(k)$ real and $r(k) = r(-k)^{\rm T}$. It follows that $\cH{0}$ is class
BDI and $\cR{0}$ is class AI. We have to show that $i_1$ is surjective. Hereto
we consider the example
\begin{align}
  r(k)&=\tau_3\cos k+\tau_1\sin k,\ \ 0 \le k \le 2 \pi,
  \label{eq:a2rpm}
\end{align}
which is nontrivial, when seen as a loop in $\cH{0}$. 

\section{Increasing the unit cell size in class AII$^{\mathcal{R}_{-}}$ by an odd number}\label{app:folding}

In this section we show how to increase the size of the unit call for a
Hamiltonian $H(k)$ with anticommuting $\mathcal{R}$ and $\mathcal{RT}= K$
symmetries by a factor $2 n + 1$, and show that this procedure does not change
the topological invariant (\ref{eq:A2inv}). 

We define the Hamiltonian $\tilde H(k)$ on the smaller Brillouin zone
$k\in[-\pi/(2n+1),\pi/(2n+1)]$ as
\begin{align}
  \tilde H(k)&=R^T(k)\left(\sum_{\substack{\oplus\\l=-n}}^n
  H\left(k+\frac{2l\pi}{2n+1}\right)\right) R(k),
  \label{eq:2n1folding}
\end{align}
where the orthogonal matrix $R(k)$ is introduced to enforce the $2\pi/(2n+1)$
periodicity of the folded Hamiltonian $\tilde H(k)$ while keeping $\tilde H(k)$
real,
\begin{align}
  R(k)&=\sum_{l=-n}^n e^{inlk}\vert v_l\rangle\langle v_l\vert,\\
  \vert v_l\rangle&=\sum_{m=-n}^n e^{imlk} \vert e_m\rangle,
  \label{eq:Rkdef}
\end{align}
$\vert v_l\rangle$ are eigenvectors of a circular permutation. The above
Hamiltonian satisfies same symmetry relations as $H(k)$ with $\mathcal{RT}=K$
and ${\mathcal{\tilde R}}=O\otimes\mathcal{R}$ with $O_{ij}=\delta_{i,-j}$.

The folded Hamiltonian has the negative-eigenvalue eigenvectors
\begin{align}
  \vert\psi_{la}\rangle&=R(-k)\vert e_l\rangle\otimes\Big\vert
  u^-_a\left(k+\frac{2\pi l}{2n+1}\right)\Big\rangle,
  \label{eq:eigenvectfold}
\end{align}
where $\vert u^-_a(k)\rangle$ are Bloch wave functions with negative energies.
The matrix $\tilde w_{ij,ab}$ from Eq.\ (\ref{eq:A2inv}) for the Hamiltonian
$\tilde H(k)$ reads
\begin{align}
  \label{eq:wijab}
  \tilde w_{ij,ab}(k)=&\langle\psi_{ia}(-k)\vert\tilde{\mathcal{R}}\vert\psi_{ja}(k)\rangle^*\\
  =&\left( R(k)O R(-k) \right)_{ij}w_{ab}\left( k+\frac{2\pi j}{2n+1} \right)\nonumber\\
  =&\left( R(k)O R(-k) \right)_{ij}\nonumber\\
  &\times\Big\langle u_a^-\left(-k+\frac{2\pi i}{2n+1}\right)\Big\vert\mathcal{R}\Big\vert
  u_b^-\left(k+\frac{2\pi j}{2n+1}\right)\Big\rangle.\nonumber
\end{align}
Since $w_{ab}(k)$ is an orthogonal matrix we have $\det w_{ab}(k)=1$. For
$k=0,\ \pi/(2n+1)$, the block on the diagonal is $w_{ab}(0)$ and $w_{ab}(\pi)$,
respectively, thus $\mbox{Pf}[\tilde w(0)]=\mbox{Pf}[w(0)]$ and
$\mbox{Pf}[\tilde w(\frac{\pi}{2n+1})]=\mbox{Pf}[w(\pi)]$. 

\section{Numerical simulations of second descendant $\bZ_2$
phase}\label{sec:simulations}

\begin{figure}
\includegraphics[width=0.4\columnwidth]{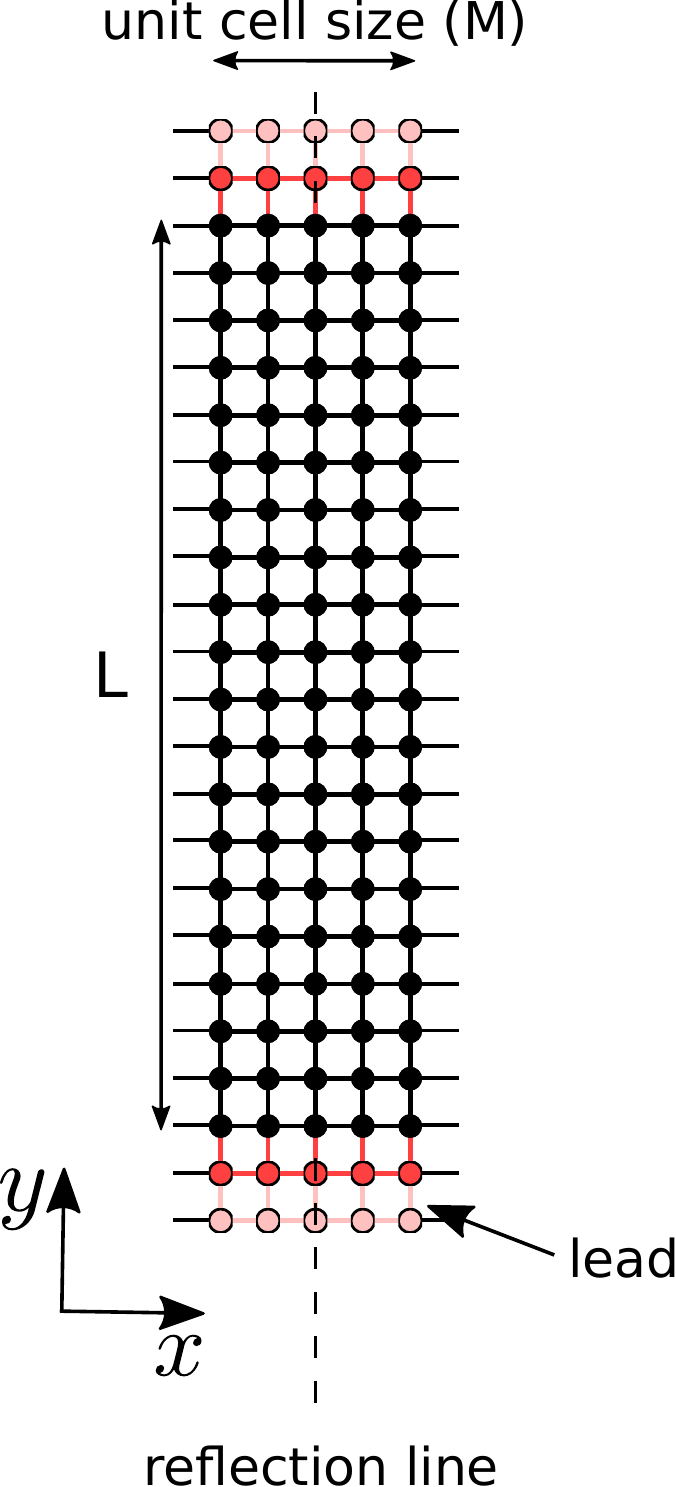}
\caption{\label{fig:CIIsetup} Schematic illustration of the transport setup we
consider. We consider two leads (red) that run along the $y$-direction,
attached to the system (black). Twisted periodic boundary conditions
(corresponding to the wavenumber $k_x$) are applied in the $x$-direction, with
unit-cell size $M$. The whole system (together with the leads) is
reflection symmetric with reflection line parallel to the $y$-axis. Each site
(point) contains eight orbitals. The system size $L$ in the $y$-direction is
chosen large enough that $r(k_x)$ is unitary.}
\end{figure}

In this section we show how to carry out the analysis of Sec.~\ref{sec:Z2}
numerically for a concrete example. We consider a 2d model with eight-band
Hamiltonian in class CII$^{\ccR{--}}$ for which it was previously argued that it
exhibits a subtle instability in the presence of
disorder.~\cite{PhysRevB.88.075142} After performing Fourier transform to 
real space, assuming the lattice to be square, the Hamiltonian reads
\begin{align}
  \label{eq:Hsys}
  H&=\sum_{x_i,x_j,y}h(k_x,x_i,x_j,y)\ket{x_i,y}\bra{x_j,y}\\
  &+\frac{1}{2}(\tau_3+i\mu_1\tau_1\sigma_2)\ket{x_i,y}\bra{x_i,y+1}+\mbox{h.c.},\nonumber
\end{align}
with the matrix elements given by
\begin{align}
  \label{eq:hxixj}
  h(x_i,x_j)&=(m\tau_3+\Delta(\vert
  x\vert)\tau_2\sigma_2)\delta_{x_i,x_j}\\
  &+\frac{1}{2}(\tau_3+i\tau_1\sigma_1)\delta_{x_i,x_j-1}+\frac{1}{2}(\tau_3-i\tau_1\sigma_1)\delta_{x_i,x_j+1}\nonumber
\end{align}
with the symmetries defined as $\mathcal{T}=\sigma_2K$,
$\mathcal{P}=\tau_1\sigma_2K$ and $\mathcal{R}=\sigma_2$. For numerical
simulations we assume the unit-cell size to be $M$ in the $x$-direction and
express the above Hamiltonian in the following basis states

\begin{align}
  \ket{k_x,\tilde x_i,y}&=\sqrt{\frac{M}{2\pi}}\sum_{m=-\infty}^me^{-ik_xmM}\ket{\tilde x_i+mM,y},
  \label{eq:periodic_basis}
\end{align}
which are $2\eta$-periodic functions of $k_x$ with $\eta=\pi/M$. We assume
$\Delta(x)$ to be a spatially non-correlated gaussian variable, with mean
$\bar\Delta$ and standard deviation $\delta$. Additionally, we take the origin
of the coordinate system to be on the reflection line, which assures that the
$\Delta(\vert x\vert)$ term is reflection symmetric. A non-zero value of
$\bar\Delta$ shifts the position of the two Dirac cones away from the mirror
line.~\cite{PhysRevB.88.075142} Each lattice site depicted on
Fig.~\ref{fig:CIIsetup} contains eight orbitals defining the space on which the
Pauli matrices $\mu_i$, $\tau_i$ and $\sigma_i$ act. Physically, $\mu_i$,
$\tau_i$ and $\sigma_i$ can be treated as particle-hole, one-half spin and
pseudospin degrees of freedom, respectively.~\cite{PhysRevB.88.075142} The above
Hamiltonian is in the topologically non-trivial phase for $\vert
m\vert<1+\sqrt{1-\bar\Delta^2}$.

\begin{figure}
\includegraphics[width=0.7\columnwidth]{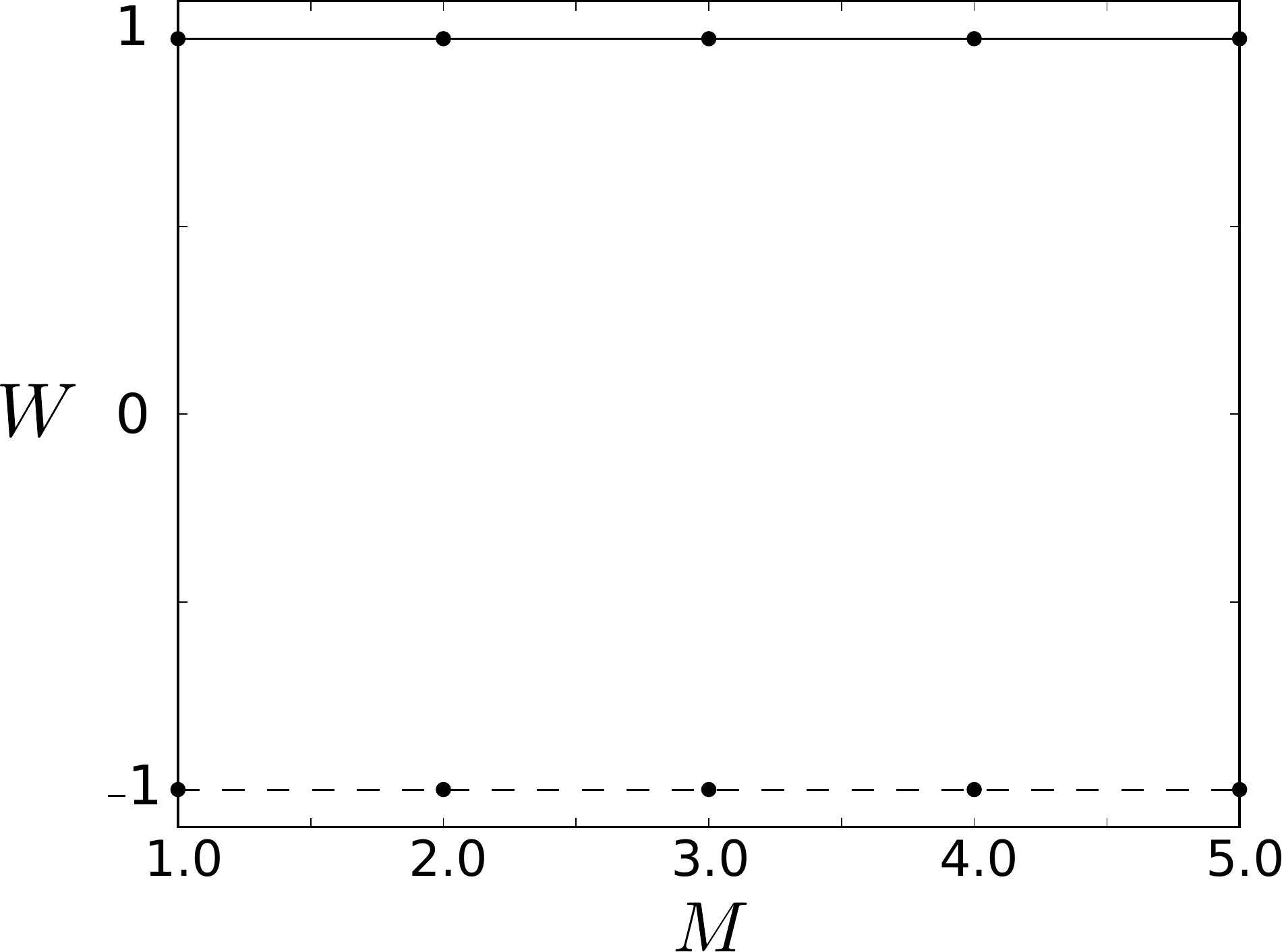}
\caption{\label{fig:Z2inv} The dependence of topological invariant
(\ref{eq:A2inv}) on the unit-cell size $M$. The solid line is for $m=3$
(topologically trivial), whereas the dashed line is for $m=1$ (topologically
non-trivial). The remaining parameters take the values $L=20$, $\bar\Delta=0.1$
and $\delta=0.03$.}
\end{figure}

We first numerically calculate the topological invariant defined by
Eq.~(\ref{eq:A2inv}). To this end we consider the setup depicted in
Fig.~\ref{fig:CIIsetup}, where two leads with a finite width are attached to the
system. For the Hamiltonian of the leads we take
\begin{align}
  \label{eq:Hlead}
  H_L=&\sum_{x,y}\left( m\tau_3\vert x,y\rangle\langle x,y\vert +\frac{1}{2}\tau_3\vert
  x,y\rangle\langle x+1,y\vert+\mbox{h.c.}\right),
\end{align}
which also belongs to the class CII$^{\ccR{--}}$. For the above choice of the
Hamiltonian, the number of propagating modes at $\varepsilon=0$ is equal to the
total number of orbitals within the unit-cell. We use software package
Kwant~\cite{groth2014} to numerically calculate the reflection matrix from the
upper lead at $\varepsilon=0$. Next, using Eq.~(\ref{eq:Hdchiral}) we obtain the
1d Hamiltonian $H(k_x)$ in class AII$^{\ccR{-}}$ with time-reversal symmetry
given by Eq.~(\ref{eq:Hd1TP}). After applying a unitary transformation to the basis
where $H(k_x)$ is real, we perform an exact diagonalization for each
$k_x\in[0,\pi]$. In order to apply Eq.~(\ref{eq:A2inv}), we need a continuous
basis of the negative energy manifold, $\vert u^-_a(k_x)\rangle$, which we find
by projecting $\vert u^-_a(k_x-\delta k_x)\rangle$ onto the negative energy
space at $k_x$ followed by orthonormalization procedure, where $\delta k_x$ is
the step size. Figure~\ref{fig:Z2inv} shows that topological invariant remains
unchanged when disorder is considered and unit-cell size is being varied, in
accordance with analysis from Sec.~\ref{sec:Z2}.

\begin{figure}
\includegraphics[width=\columnwidth]{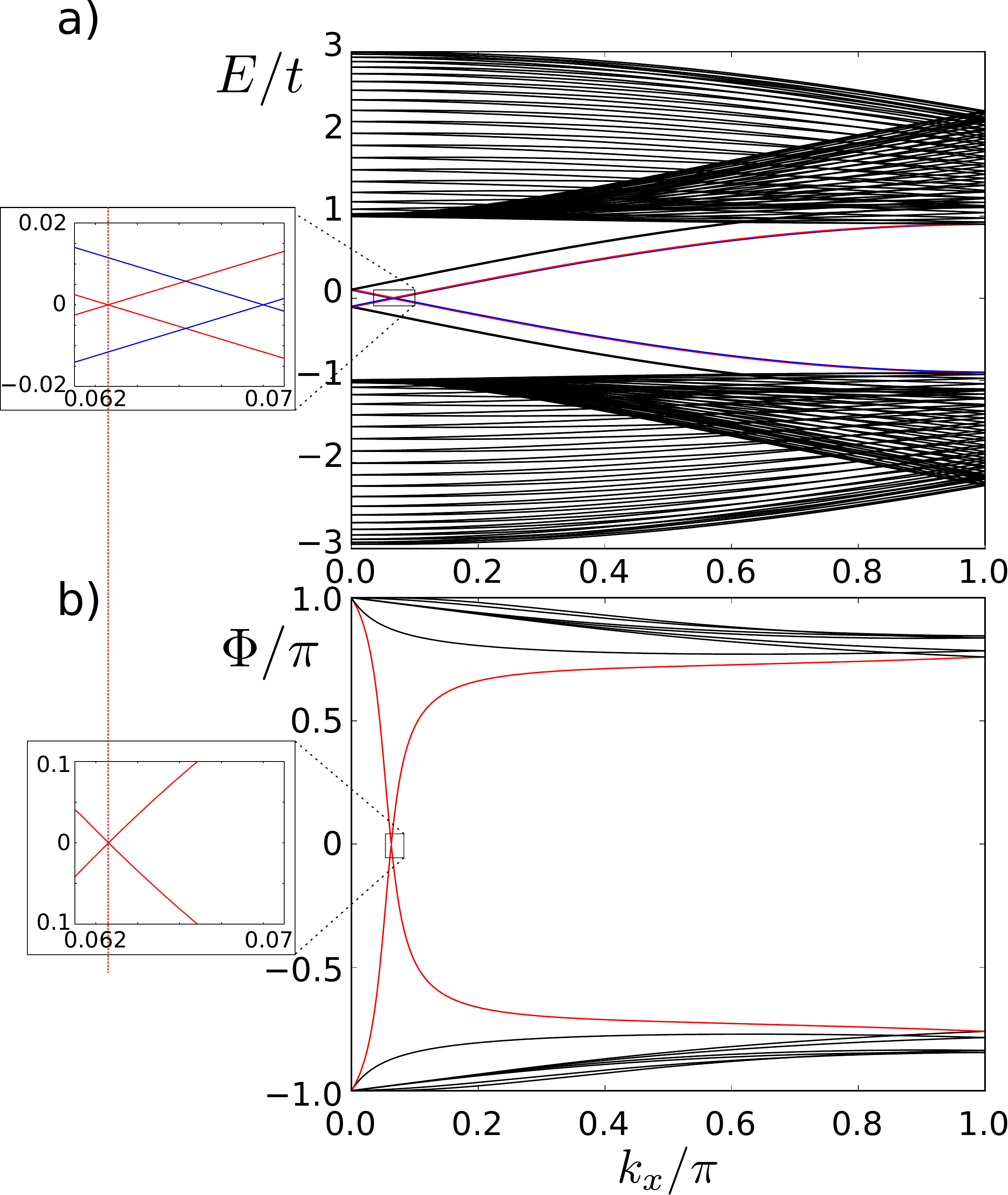}
\caption{\label{fig:CIIplots} a) Dispersion over half of the Brillouin-Zone
$k_x\in[0,\pi]$ for the system depicted in Fig.~\ref{fig:CIIsetup}. The inset
shows zoomed-in spectra that contains two Dirac-like dispersions, the red (blue)
states being localized on the top (bottom) edge of the sample. b) Phases $\Phi$
of the eigenvalues of the unitary operator $r(k_x)r^\prime(k_x)$, where $r(k_x)$
is the reflection matrix from the upper lead. The inset show that the position
of the red Dirac cone matches the position where the phase is zero. The
remaining parameters take the values $L=20$, $M=2$, $m=1$, $\bar\Delta=0.1$ and
$\delta=0.03$.}
\end{figure}

We next show how to numerically determine, from the reflection matrix, the
presence of the edge states in the $x$-direction and compare these results with
the exact diagonalization of the full two-dimensional
Hamiltonian~(\ref{eq:Hsys}). We show the results for the case of doubled
unit-cell ($M=2$) but all the conclusions that follow are valid for arbitrary
$M$. Figure~\ref{fig:CIIplots}a shows the band structure for $k_x\in[0,\pi]$.
There are four Dirac cones, two at the upper edge and two at two at the lower
edge. Since we take $\bar\Delta\neq0$, the Dirac cones of each edge are moved
away from $k_x=0$ line.

\begin{figure}
\includegraphics[width=0.8\columnwidth]{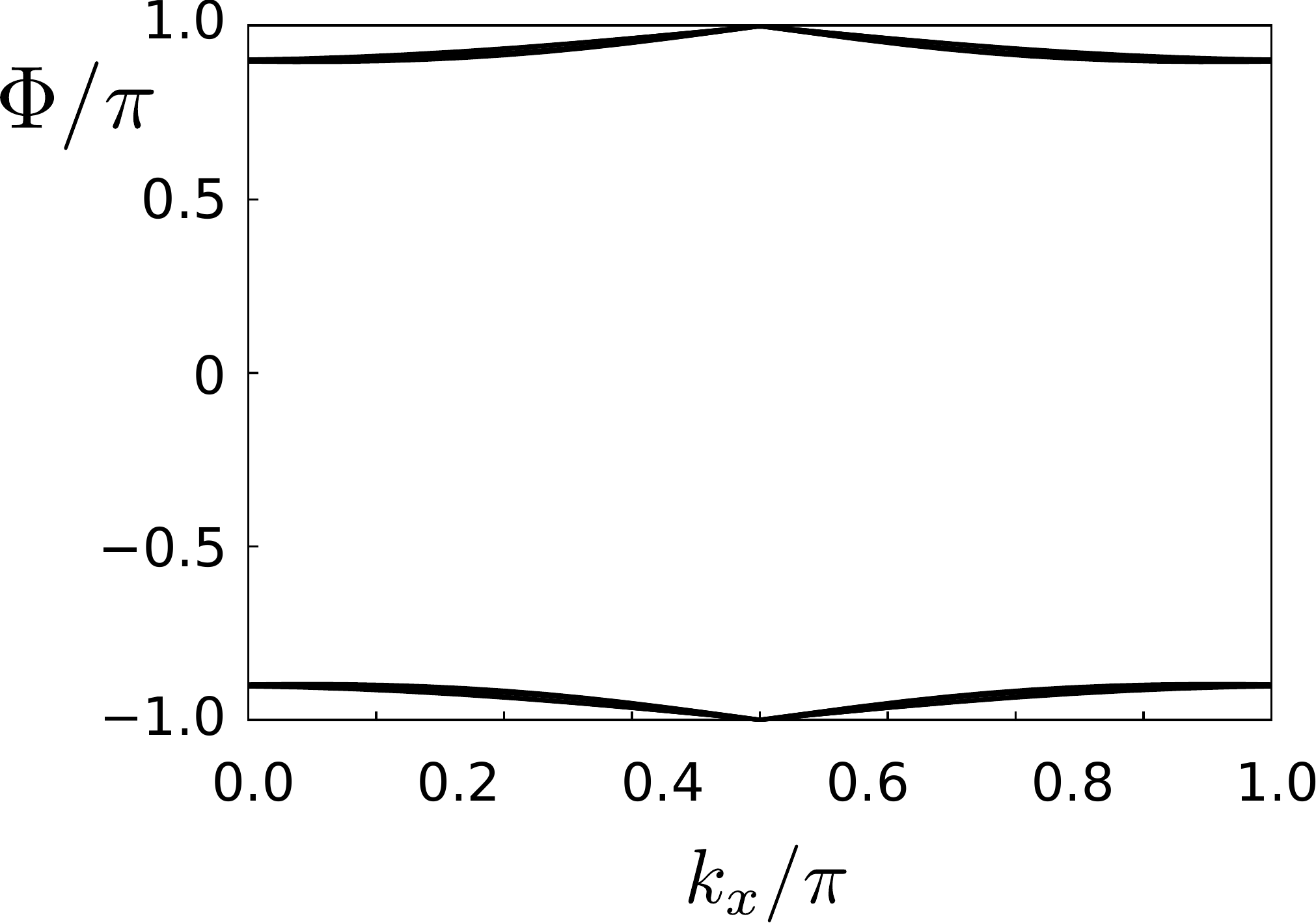}
\caption{\label{fig:CIIPHItrivial} Phases $\Phi$ of the eigenvalues of the
unitary operator $r(k_x)r^\prime(k_x)$, where $r(k_x)$ is the reflection matrix
from the system in the trivial phase. The remaining parameters take the values
$L=20$, $M=2$, $m=3$, $\bar\Delta=0.1$ and $\delta=0.03$.}
\end{figure}

In order to apply Eq.~(\ref{eq:boundstates}) we consider leads that are infinite
in the $x$-direction so that $k_x$ is conserved. For $r^\prime(k_x)$ we take
a reflection matrix that transforms the incoming to the outgoing modes without
affecting the mode wave-function within the unit-cell. The reflection matrix
$r(k_x)$ we obtain numerically by considering scattering from the upper lead
for each $k_x\in[0,\pi]$. Finally, we transform the unitary operator
$r(k_x)r^\prime(k_x)$ to the basis that is $k_x$-independent and perform the
exact diagonalization.  Figure~\ref{fig:CIIplots}b, shows the phases
$\Phi$ of the obtained eigenvalues.  By comparing panels a) and b) we conclude
that the position of the Dirac cone (on the upper edge) and the appearance of
the eigenvalue equal to one (i.e. $\Phi$ equal to zero) of the product
$r(k_x)r^\prime(k_x)$ occur at the same $k_x$ value. As explained in the main
text, the presence of a single (or in general odd number) value of $k_x$ for
which $\Phi=0$ is guaranteed by the topological invariant $W$ being non-trivial.
For the sake of completeness, in Fig.~\ref{fig:CIIPHItrivial} we show that there
is no value of $k_x$ for which $\Phi=0$ (or an even number of such values,
depending on the termination $r^\prime(k_x)$) if the topological invariant $W$
is trivial.

We also explicitly considered the perturbation which was previously argued to
gap out the edge states the model (\ref{eq:Hsys}), see
Ref.~\onlinecite{PhysRevB.88.075142},
\begin{align}
  \delta h(x_i,x_j) =&- \frac{4c}{\pi}\mu_3\tau_2\sigma_1\sin[\eta(x_i+x_j)]\nonumber\\
  &\times\frac{\cos[\delta_1(x_i-x_j)]}{x_i-x_j},
  \label{eq:deltaH1}
\end{align}
where $\delta_1$ is a parameter; Its value should be chosen close to the
position of an edge state Dirac cone. Although this perturbation breaks
translation symmetry on the level of a single unit cell, the
translation-symmetry breaking can be chosen to be commensurate with the
underlying lattice by choosing rational $\eta = \pi/M$, so that it can be
included in a Bloch Hamiltonian $H(k)$ for a unit cell size $M$. We confirmed
that the perturbation (\ref{eq:deltaH1}) opens a gap in the edge-state spectrum
without closing the bulk gap for $c \gtrsim 1$. A gap in the edge-state
spectrum is also opened up for $c \gtrsim 1$ if the perturbation is added
near the sample edge only. However, the perturbation (\ref{eq:deltaH1}) is not a
local perturbation; It has long-range hopping, with a hopping amplitude decaying
inversely proportional to distance. Hence, the corresponding Bloch Hamiltonian
is not a continuous function of $k$, so that the topological classification,
which assumed a continuous $k$ dependence, does not apply. To remain within the
paradigm of the topological classification, we replace Eq.\ (\ref{eq:deltaH1})
by a perturbation with a short-range hopping term, while otherwise preserving
the matrix structure,
\begin{align}
  \delta h(x_i,x_j) =&-\frac{2c}{\pi}\mu_3\tau_2\sigma_1\sin[\eta(x_i+x_j)]\nonumber\\
  &\times\cos[\delta_1(x_i-x_j)]e^{-a\eta\vert x_i-x_j\vert},
\end{align}
with $a$ a numerical constant of order unity. This gives a continuous Bloch
Hamiltonian $H(k)$. For this Hamiltonian no gap in the edge-state spectrum was
found for arbitrary strength of the perturbation if the perturbation is added
near the sample edge only. If the perturbation is added uniformly in space, the
opening of a gap in the surface state spectrum is preceded by the closing of the
bulk gap, signalling a topological phase transition.

\bibliography{refs}
\end{document}